# Observed circulation trends in boreal summer linked to two spatially distinct teleconnection patterns

Tamara Happé,[a] Chiem van Straaten,[a,b] Raed Hamed,[a] Fabio D'Andrea[c], Dim Coumou[a,b].

[a] *Institute for Environmental Studies at Vrije Universiteit Amsterdam, Amsterdam, the Netherlands.*

[b] *Royal Dutch Meteorological Institute (KNMI), de Bilt, the Netherlands.*

[c] *Laboratoire de Météorologie Dynamique/IPSL, ENS, PSL Research University, École Polytechnique, Institut Polytechnique de Paris, Sorbonne Université, CNRS, Paris, France.*

*Corresponding author*: Tamara Happé, t.happe@vu.nl

ABSTRACT

Various regions in the Northern Hemispheric midlatitudes have seen pronounced trends in upper-atmosphere summer circulation and surface temperature extremes over recent decades (since 1979). Several of these regional trends lie outside the range of historic CMIP6 model simulations, and they might constitute a joined dynamic response that is missed by climate models. Here, we examine if the regional trends in circulation are indeed part of a coherent circumglobal wave pattern. Using ERA5 reanalysis data and CMIP6 historical simulations, we find that the observed upper-atmospheric circulation trends consist of at least two separate regional signatures: a US-Atlantic and a Eurasian trend pattern. The circulation trend can explain on average 15% and 26% of the observed regional temperature trends in the US-Atlantic and Eurasian regions respectively. The circulation trend in the CMIP6 multi-model-mean does not resemble the observed trend pattern and is much weaker overall. Some individual CMIP6 models do show a resemblance to the observed pattern in ERA5, although still weak with a maximum pattern correlation of 0.47. The pattern correlation is higher for the two individual regions (US-Atlantic and Eurasia), reaching a maximum of 0.69 and 0.78. We show that both regional wave patterns in ERA5 are associated with distinct sea surface temperature and outgoing longwave radiation anomalies in the three weeks leading up to the atmospheric configuration, resembling known teleconnection patterns. CMIP6 models appear to lack these tropical-extra-tropical teleconnections. Our findings highlight the limitations of CMIP6 models in reproducing teleconnections and their associated regional imprint, creating deep uncertainty for regional climate projections on decadal to multi-decadal timescales.

SIGNIFICANCE STATEMENT

With climate change several regions around the world are experiencing trends in both atmospheric- and in surface variables, such as temperature. Some of these trends in observational datasets lay outside of the range of the trends as simulated in state-of-the-art climate models. In this research we show that there are two separate regions in which the regional atmospheric trends behave consistent with each other (the US-Atlantic and Eurasian sectors). These atmospheric circulation trends can explain up to 30% of the observed regional surface temperature trends. Critically, the state-of-the-art climate models do not have the same circulation trend patterns and physical pathways as found in the observations. This can be problematic for regional climate projections on decadal to multi-decadal timescales.

## 1. Introduction

Trend analysis on recent decades has revealed that various regions in the Northern Hemispheric midlatitudes experience a marked set of changes in surface temperature and atmospheric circulation in summer (e.g. Teng et al., 2022). For example, temperatures and heat extremes in Europe are rising faster than other regions around the globe (Rousi et al. 2022; Patterson 2023; Singh et al. 2023; Vautard, et al. 2023) (Fig. A1). Heat extremes have devastating impacts on human health, ecosystems, agriculture, and the energy sector (Ballester et al. 2023; Barriopedro et al. 2023). In contrast to the excess warming in Europe, the eastern US has seen hardly any warming in temperature extremes, i.e. much less than other midlatitude regions (Singh et al. 2023). Both western European and US surface temperature changes have been partly attributed to circulation changes (Singh et al. 2023). Other notable changes in circulation in the Northern Hemisphere are positive trends over western Russia, central China, western US, and Greenland and negative trends over the Caspian Sea and Alaska (e.g. Teng et al., 2022).

Several of the observed regional trends are not captured by state of the art climate models (CMIP6). For example, climate models fail to capture the western European accelerated temperature trend as seen in observations (Van Oldenborgh et al. 2022; Vautard et al. 2023). Part of this mismatch has been attributed to changes in atmospheric circulation, through an increase in frequency of the extended subtropical ridge (Vautard et al. 2023). Similarly, the increase in deep depressions in the North Atlantic Ocean is also outside of the range of climate model simulations over the historic period (D’Andrea et al. 2024). Furthermore, the increase in Greenland Blockings are not captured by state of the art climate models, potentially due to missing teleconnections to sea surface temperatures (SSTs), Sea Ice Coverage, and/or diminished snow cover (Preece et al., 2023; Maddison et al. 2024). Critically, it is not yet understood whether those circulation trends are a forced response to greenhouse gases, or due to internal variability in the climate systems. Neither is it known what the underlying reason is for the mismatch with climate models.

Recent literature explains the boreal summer trend pattern in upper-level-atmospheric circulation as one circumglobal wave (Sun et al. 2022; Teng et al. 2022), however, a full physical understanding is still lacking. Kim et al. (2022) showed that the circumglobal teleconnection pattern from Branstator (2002) is not linked to the heat extremes in the Eurasian sector, indicating this circumglobal teleconnection is not responsible for the observed temperature trends in this region. This example highlights the lack of consensus on whether there is indeed one circumglobal wave that causes the observed trend pattern. Furthermore, there are no clear physical pathways defined that could explain the circumglobal observed trend pattern. Some have linked the changes in circulation to SST patterns. For example, Teng et al. (2022) find that the trend pattern is linked to multi-decadal variability in SST patterns, notably the Pacific Decadal Oscillation and Atlantic Multidecadal Variability. Similarly, Sun et al. (2022) argue that the observed wave is coming from SST trends in the Western North Pacific (WNP), and associated changes in upper-level convergence. Evidently, this underscores the broader lack of consensus – not only on whether the observed atmospheric circulation trend is due to one increasingly prevalent circumglobal wave, but also on their underlying physical drivers. Understanding these two questions is the first step in investigating the discrepancies between the observations and the current state of the art climate models (CMIP6), and is thus vital for short term predictions as well as long term climate projections

Understanding the underlying processes behind atmospheric circulation changes is complex due to (multi-)decadal variability, interannual variability, as well as intra-annual variability determining both low- and high-frequency processes (e.g. Shaw et al., 2024, preprint). For example, there's known interannual variability of the Northern Hemisphere to ENSO (Ding et al. 2011; Jong et al. 2020), as we also show in Fig. A2 in the appendix. In this research we use multiple decomposition methods on different spatial- and temporal-timescales. In order to decompose the circulation trends further, we will use a decomposition method called rank-anomalies (Röthlisberger et al. 2020), which allows us to assess the different parts of the distribution separately. We combine different statistical methods to understand which parts of the trend-pattern are connected, what potential underlying drivers are of the observed trends, and why climate models fail to capture those.

## 2. Methods

In this research, we use both observations and climate model simulations (section 2.1) to investigate the trends, potential drivers, and mismatch between observations and models. We use different decomposition methods, including preprocessing and rank anomalies, understand the underlying patterns leading to the observed atmospheric circulation trends (section 2.2). By analyzing trends across different regions and timescales, from sub-seasonal to decadal to multi-decadal, we identify regions that exhibit similar and coherent behavior. We then shed light on the potential processes underlying these coherent wave-patterns in section 2.3, by looking at different climate variables preceding the atmospheric circulation anomalies. Lastly, we investigate to which extent the observed circulation trends can explain the observed surface temperature trends in section 2.4. We also provide an overview of the methodological framework at the end of this section (Fig. 2).

## 2.1 Data

In this research, we use ERA5 reanalysis data (Hersbach et al., 2019) and a subset of CMIP6 historical climate model simulations data (see table A1 for a list of the models). We use summertime (JJA) daily data available from 1940-2023 for ERA5 and 1950-2014 for CMIP6. For comparison reasons, we re-gridded the CMIP6 data to the ERA5 grid, to a 0.25 degrees spatial resolution, using a first order conservative interpolation using Climate Data Operators (CDO) (Schulzweida, 2023). The variables of interest are streamfunction at 250hPa (STREAM250) (35ºN,70ºN;-180ºE,180ºE), Surface Air Temperature (T2M) (35ºN,70ºN;-180ºE,180ºE), Sea Surface Temperatures (SSTs) (-40N,90N;-180E,180E), and Outgoing Longwave Radiation (OLR) (-20ºN,30ºN;-180ºE,180ºE). We calculated the streamfunction from the u- and v-wind fields using CDO (Schulzweida, 2023). Lastly, we use the Global Mean Surface Temperature (GMST) from NASA, which is smoothened with a 5-year LOWESS regression to filter out short-term fluctuations.

## 2.2 Decomposition methods

To determine which regional changes in the midlatitudes are connected to each other, and which processes are behind these atmospheric circulation changes, we use a variety of decomposition methods. First, we use multiple preprocessing steps, where we remove just the seasonality or also the hemispheric interannual variability. The data is de-seasonalized by removing the long-term week-of-year averages over the respective spatial domain. The hemispheric interannual variability in the streamfunction is removed in most parts of our research as well (Figures 3-5), as the entire northern hemisphere has a year-to-year fluctuation in positive and negative anomalies (see Fig. A2 for more details). Therefore, to analyze the streamfunction anomalies with respect to this hemispheric interannual-variability, we remove this fluctuation. We do this by removing the large-scale summer mean from each time step, obtaining the grid level anomalies relative to the hemispheric mean. This is essential because we correlate time series between different hotspots in the northern hemisphere. Without removing this hemispheric interannual variability, shared low-frequency fluctuations could lead to misleading ('spurious') correlations between different hotspot regions.

We further decompose the atmospheric circulation (streamfunction at 250 hPa) data using rank anomalies (Röthlisberger et al. 2020). Rank anomalies are calculated following two steps: 1) the days in each season are ranked from low to high based on their streamfunction value, given the 92 days in each summer and 2) the anomalies of each rank-day are calculated with respect to the mean of all the same-rank days throughout all the years (Fig. 1). The rank-anomaly of a day then provides information on the anomaly with respect to a normal low or high value day, as visualized in Figure 1. We then split the season into three parts: low (23 days), bulk (46 days), and high (23 days). This allows us to assess changes in the different parts of the season individually within our trend analysis, as a positive trend in the mean could also be due to an absence of low values.

We calculate the streamfunction and temperature trends per grid point using linear regression and statistical significance is calculated using a two-sided t-test. We calculate the trend over all data points, but also calculate the streamfunction trends for the three different rank-anomaly parts (low, bulk, high). Calculating the trends for each individual part of the season allows us to assess how the different parts of the distribution are changing (see appendix Fig. A3 for further details). We define a hotspot as such, if they show strong (changing) trends over time and/or if they have been previously identified by other researchers as such. We assess how the ERA5 circulation trends compare to those in CMIP6 for the hotspots and all gridpoints in the midlatitudes. To this end, we create a rank histogram to see where the ERA5 trends in comparison to the model simulations.

To be able to determine which hotspots are connected to each other, we further decompose the streamfunction data of each hotspot by applying a 21-year rolling mean to the aggregated low-, bulk-, and high-rank anomalies resulting in the decadal trend from 1950 onwards (rolling mean is centered). Removing the decadal trend from the timeseries results in the seasonal anomalies. To assess which hotspots are connected to each other, we calculate the correlations for both seasonal anomalies and decadal trends. The decadal correlations are calculated from 1950 onwards, and the seasonal anomalies from 1979 onwards. We then calculate the Pearson-correlation between each hotspot for the decadal trend and seasonal anomalies. We determine statistical significance using a p-value threshold of 0.05, using a two-sided t-test. We use the correlation matrices to find the regions that behave coherently on both timescales. If a process is responsible for changes in multiple hotspots over time, we expect the correlations to be of equal sign in both the seasonal and decadal time series. Thus, we only assess those correlations that are both statistically significant and coherent in sign.

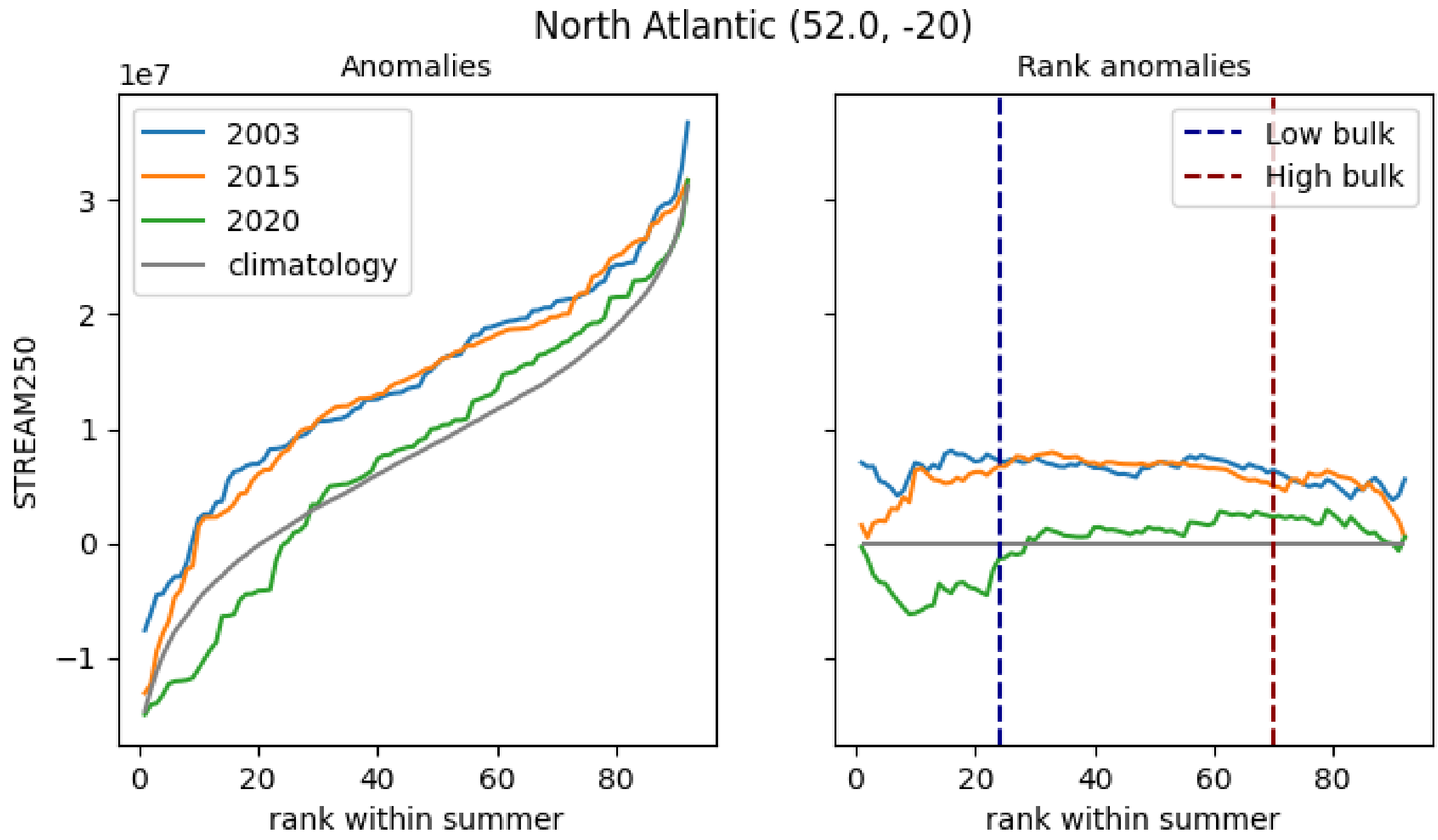

**Figure 1. Example of rank anomalies for three years, for the hotspot in the North Atlantic.** Here, three different summers are shown (2003, 2015, 2020) and the climatological values for each rank. The right panel then shows the rank anomalies, with respect to the climatology of each rank.

### 2.3 Process understanding

We use pattern matching to better understand the physical processes behind the trends and to assess differences between ERA5 and CMIP6. Pattern matching consists of calculating the spatial correlation between two fields, where the fields consist of anomalies with respect to their own spatial mean. Furthermore, we weigh the correlations with respect to the latitude using cosine.

First, we calculate the pattern correlation between the streamfunction 250hPa trend of each CMIP6 model simulation and the trend as observed in ERA5, to determine the similarity of the patterns. To this end, we calculate both the trends in CMIP6 and in ERA5 over the period 1979-2014. We assess the pattern correlation for the entire northern hemispheric midlatitudes, as well as for the individual regions found in the analysis.

Second, to better understand the processes behind the trend-patterns in the ERA5 reanalysis data, we calculate the pattern correlations between the daily streamfunction anomalies at 250 hPa data and the trend-pattern from 1979-2023, for the different regions found in the analysis. We then aggregate the correlations to a weekly timescale and use a threshold of 0.5 to select so-called ‘positive matches’ to the trend pattern. Then, we calculate the mean of different variables (SST, OLR, STREAM250, and T2M) for the positive matches, to create composite plots. We calculate the composites on different time lags to assess possible physical mechanisms leading up to the atmospheric composition of interest. Statistical significance is calculated using a two-sided t-test (Welch’s), to assess whether the anomalies of the positive matches are statistically different from the non-positive anomalies. To be able to compare with CMIP6, we create composites of the positive matches between 1950 and 2014.

Lastly, to assess whether the same physical pathways exist in the CMIP6 model simulations, we calculate the correlation time series between the model’s streamfunction 250hPa data and the ERA5 trend-pattern for the separate regions (ERA5 trend of 1979-2014). Again, we take a threshold of 0.5 to find the weeks that are positive matches and calculate the composites for the SSTs on different time lags. We then calculate the pattern correlation between the model’s SST composites and the ERA5 SST composites. We mask out the spurious and non-statistically significant regions from ERA5 and also apply latitude weighting. We furthermore plot a super-composite of all positive matches in the CMIP6 models.

**2.4 Temperature trend attribution**

To determine whether the observed circulation trends are partly responsible for the observed temperature trends, we use a simple linear regression model. We predict the daily T2M in each gridpoint using two predictors: the yearly Global Mean Surface Temperature (GMST) and the pattern correlation with the STREAM250 trend of the previously identified region of interest (Eq. 1). Using the coefficient ß2, we compute the T2Mdynamic, which is the T2M as predicted from the pattern correlation of one region (Eq. 2). Then, we calculate the linear trend of T2Mdynamic and the percentage this represents compared to the total observed temperature trend. We mask out trend values below 1E-2 from the observed trends to avoid spurious high percentages. We repeat this regression analysis for attribution for the two identified regions, and calculate the latitude-weighted mean of the explained (absolute) percentage for each region.

$$\boldsymbol{Equation\ 1} \quad \boldsymbol{T2M(t) = GMST(t) * \beta_1 + PatternCorrelation(t) * \beta_2}$$

$$\boldsymbol{Equation\ 2} \quad \boldsymbol{T2M_{dynamic}(t) = PatternCorrelation(t) * \beta_2}$$

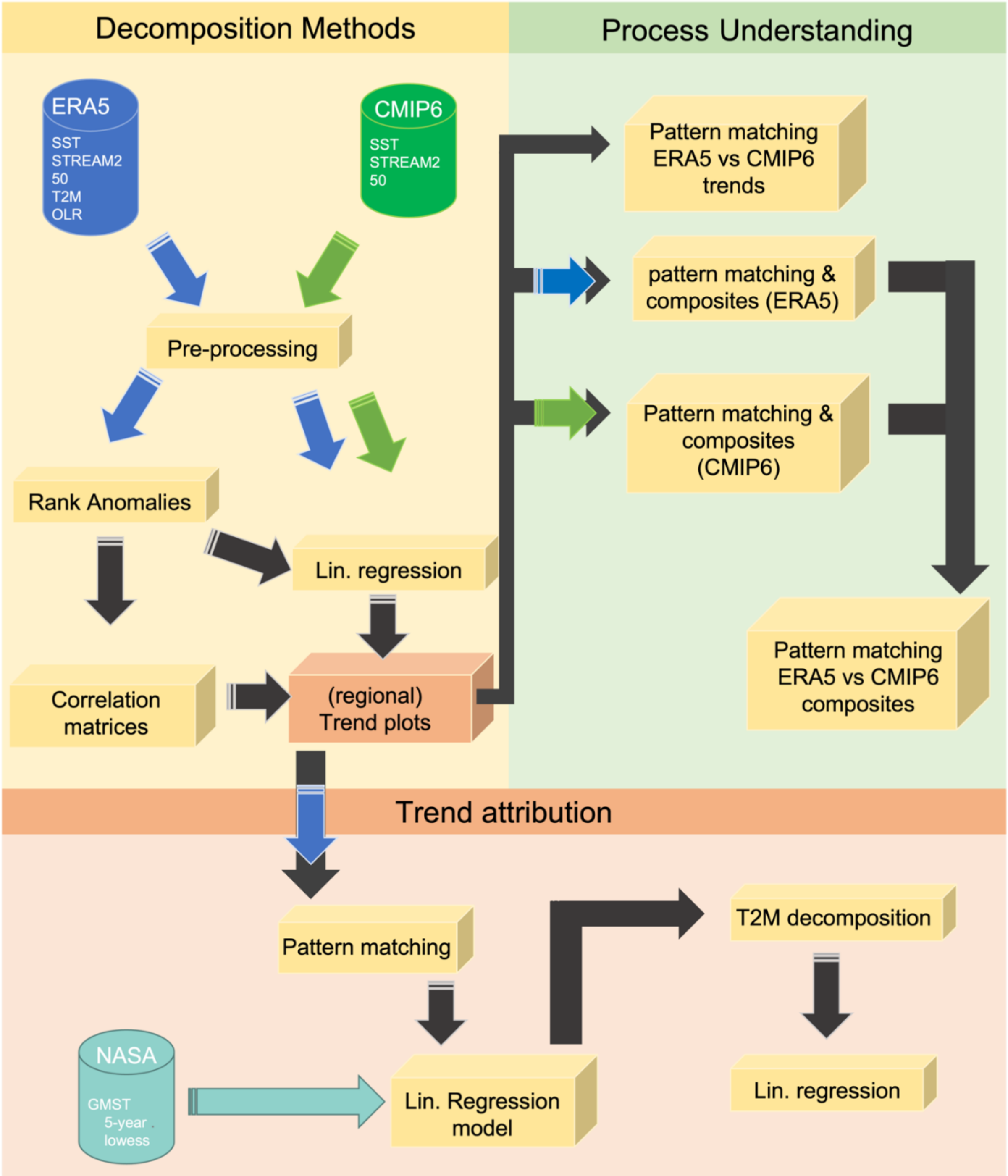

**Figure 2. Methodological overview**. A rough pipeline description of the different data and methods used in this paper. The three different boxes are further detailed in the method sections 2.2 (Decomposition methods), 2.3 (Process understanding) and 2.4 (Trend attribution).

## 3. Results

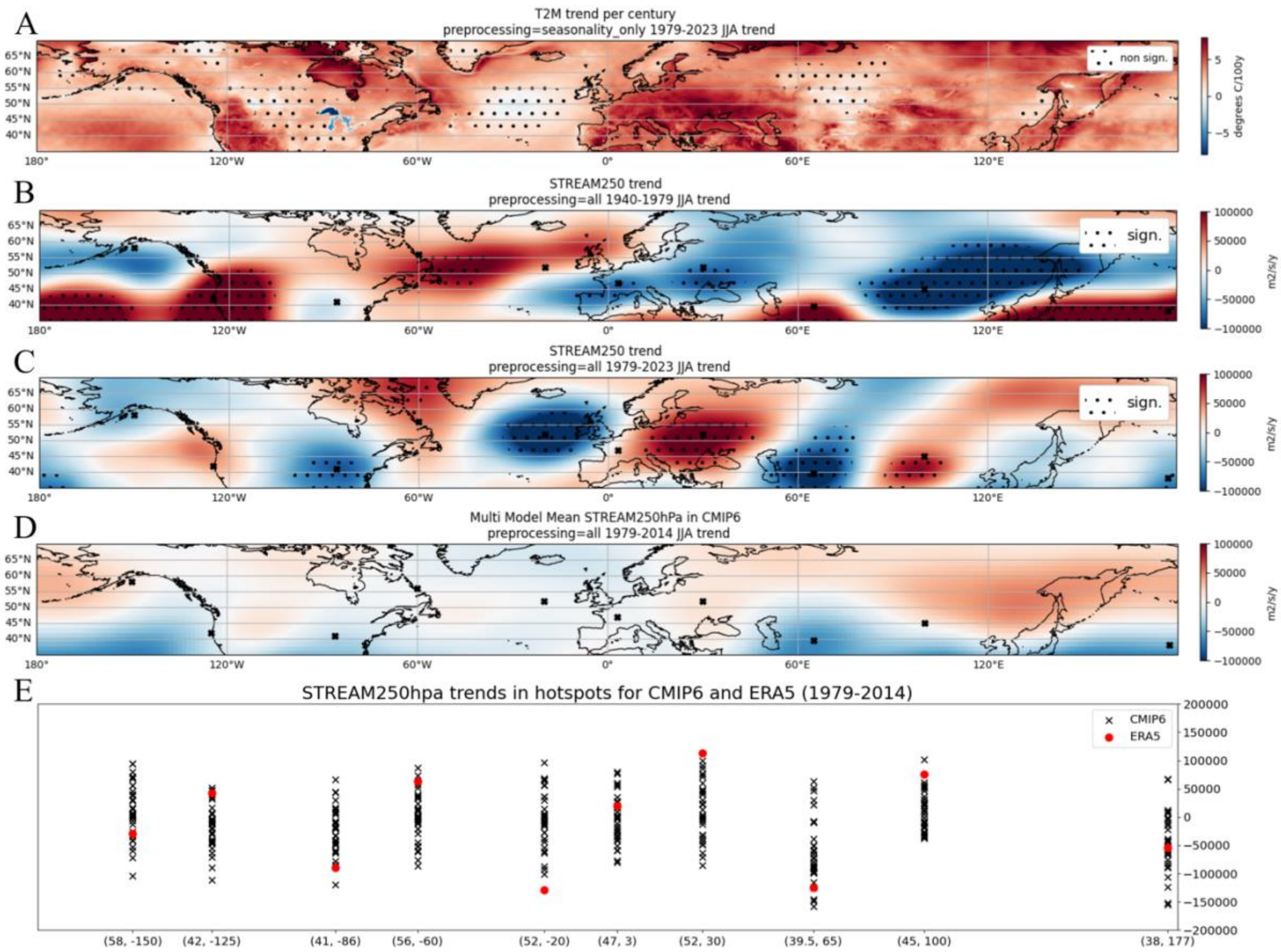

**Figure 3. Midlatitude trendplots of boreal summer with hotspots marked with X. A)** the temperature trend (per century) in degrees C (1979-2023), **B)** the yearly streamfunction at 250hPa trend in $m^2/s$ (1940-1979), **C)** the yearly streamfunction at 250hPa trend in $m^2/s$ (1979-2023), **D)** the multi-model-mean of yearly streamfunction at 250hPa trends of CMIP6 models in $m^2/s$ (1979-2014), and **E)** for each hotspot the trend in CMIP6 models (one cross per model) and the trend in ERA5 (from 1979 to 2014) for streamfunction at 250hPa. Non-statistical significant areas are stippled in panels A, B, and C, following a two-sided t-test.

### 3.1 Boreal summer trends in the northern hemisphere midlatitudes

The Northern Hemispheric midlatitudes exhibit strong surface temperature and upper-atmosphere circulation trends in boreal summer (Fig. 3a-c, Fig. B6). The surface temperature trends reflect the upper atmospheric circulation trends, with highs and lows following a wave-like trend pattern (Fig. 3a, 3c, and Fig. A2). Notably, we find statistically significant trends in the circulation in most hotspots: East US (-), Greenland (+), North Atlantic (-), East Europe (+), east of the Caspian Sea (-), and in Central China (+). Some trends already appeared before 1979 (Fig. 3b), like the US-Greenland high-low-high, while other trends only emerged after 1979, like the Eurasia high-low-high. Fig 3d shows the multi-model mean streamfunction trend at 250hPa of 33 CMIP6 historical simulations (1950-2014). The trend pattern as observed in ERA5 does not appear in the CMIP6 multi-model-mean. To further investigate the trends of the individual model simulations, we plot the trend in each hotspot in Fig 3e. It becomes apparent that in several hotspot regions, the observed trend is outside or at the top of the CMIP6 model simulation spread. Furthermore, we show that the magnitude of both positive and negative circulation trends in ERA5 are generally underestimated by the CMIP6 models (Fig B4).

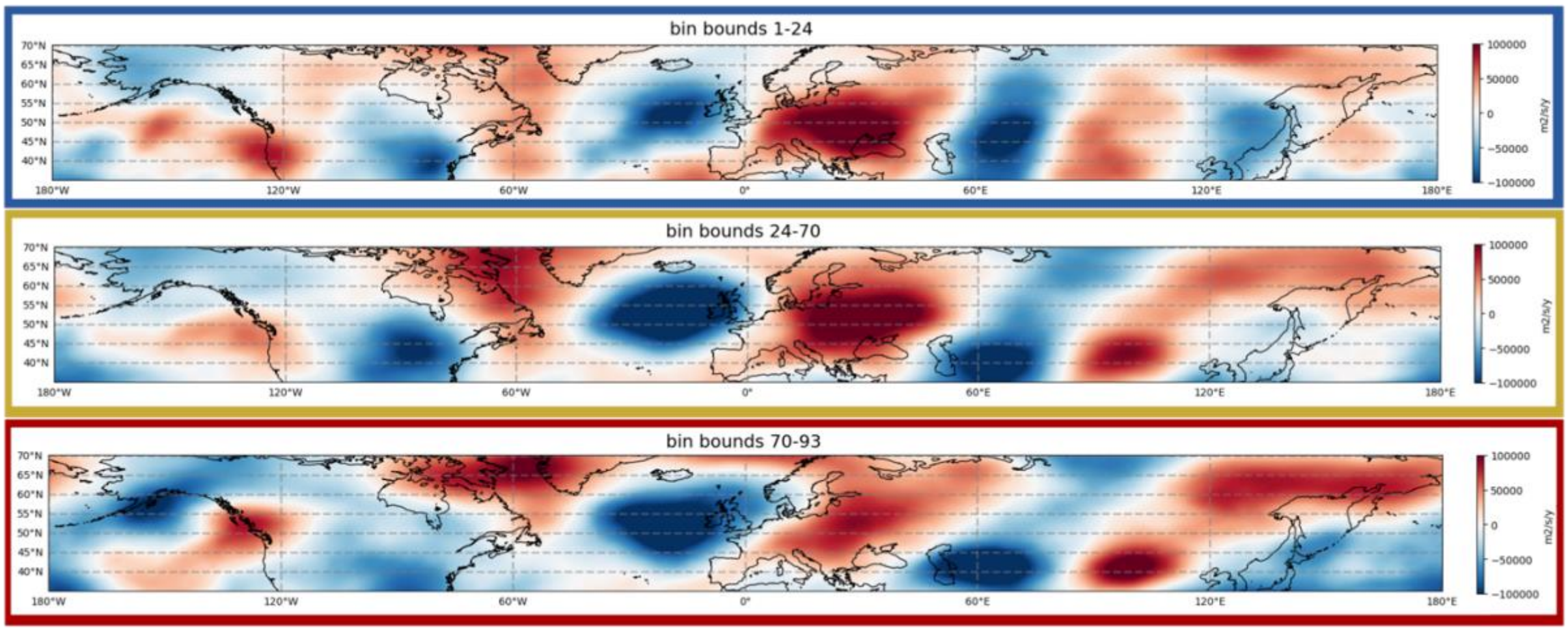

**Figure 4. Decomposition of trend pattern for the period 1979-2023.** The panels show the trend for the low (top panel), bulk (middle panel), and high (bottom panel) parts of the distribution, from top to bottom respectively. The difference between the overall trend and the trend in the different parts of the distribution are shown in the appendix (Fig. B1).

### 3.2 Trend pattern decomposition

Using rank anomalies, we calculate the trend in each grid point for the different parts of the distribution: low, bulk, and high. To the first degree, we find a similar pattern in each part of the distribution (Fig. 4). However, there are regions which show clear differences in the parts of the distribution (Fig. B1), meaning the trend is not equal throughout the distribution. For example, the southwest US shows a clear increase in the low parts of the distribution, meaning the low pressures are increasing, whereas there is no positive trend in the high pressure. The same pattern occurs in parts of central Europe. On the other hand, we see regions that see consistent changes throughout all parts of the distribution: the North Atlantic, west Russia, the Caspian Sea, and west China. It is clear that for Greenland, the high parts of the distribution are increasing the most, indicating the upper tail of the distribution is experiencing the biggest increase. Opposite of this is the Alaskan hotspot, where the high pressures are decreasing significantly. It is important to note that a positive trend does not automatically mean an increase in high pressures (and thus heat extremes), instead it is important to consider the changes in each part of the distribution separately.

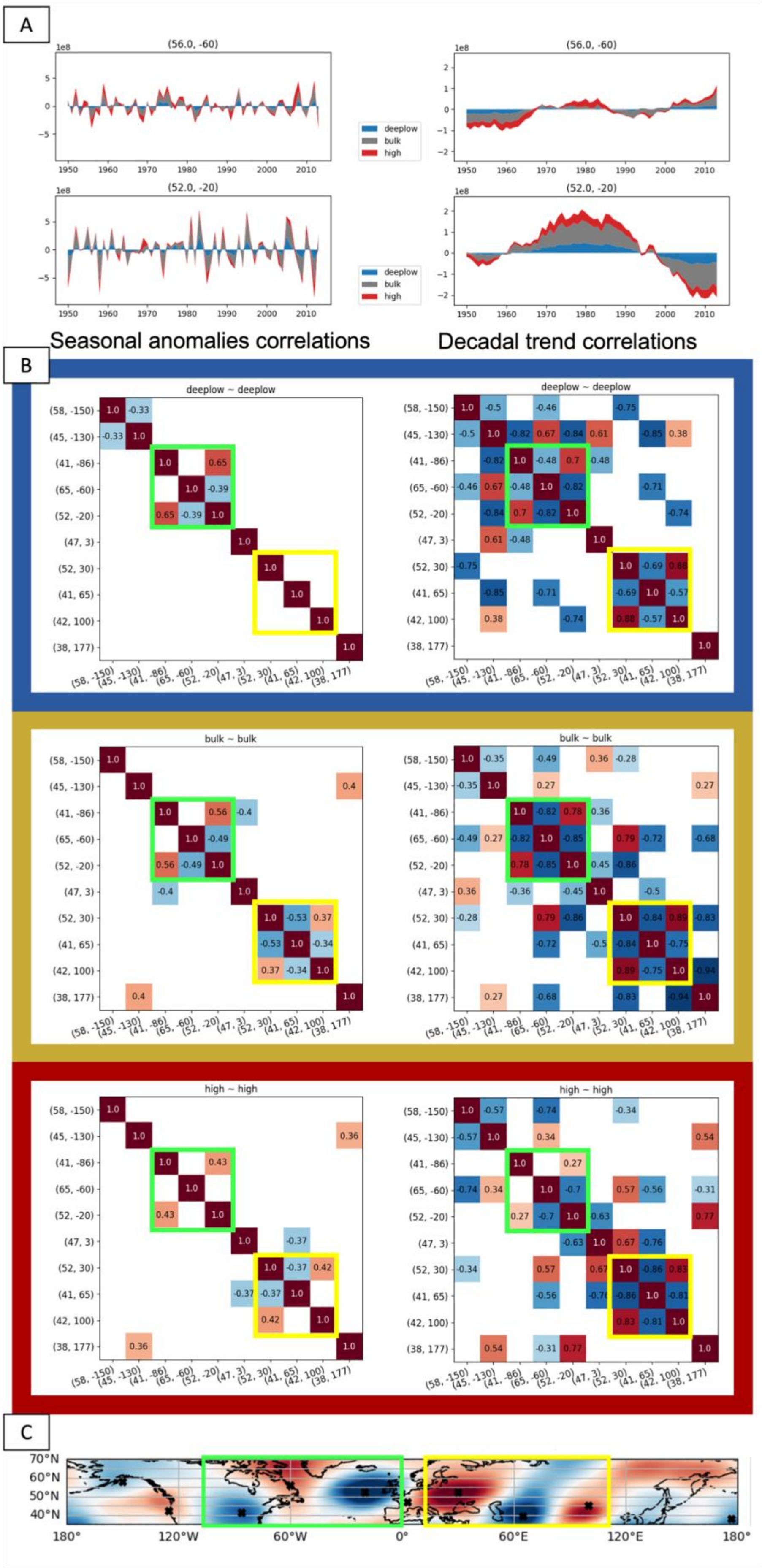
A
(56.0, -60)
(52.0, -20)
deeplow
bulk
high
Seasonal anomalies correlations
Decadal trend correlations
B
deeplow ~ deeplow
bulk ~ bulk
high ~ high
C
70°N
60°N
50°N
40°N
180°
120°W
60°W
0°
60°E
120°E
180°

**Figure 5. Rank Anomaly correlations for seasonal and decadal coherence. A) evolution plots of two example hotspots,** with the seasonal anomalies and decadal evolution for each rank anomaly part, **B) correlation matrices** for seasonal anomalies and decadal trends between hotspots for low (blue), bulk (yellow), and high (red) rank anomalies, with each hotspot on the x- and y-axis and C) trend plot with corresponding regions and hotspots. For more details please see method section 2.2.

### 3.3 Regional consistent behavior

If the trend pattern is the result of single or multiple drivers that create a circum-global wave pattern, then we expect the hotspot regions to show some co-variability on both intra-seasonal and decadal timescales. To test this, we correlate the decadal trend and seasonal anomalies of the hotspots with each other. For illustration, Fig. 5a shows for two hotspots the decadal trend (21-year rolling mean) and the seasonal anomalies (residuals of the 21-year rolling mean), for each part of the distribution as obtained with the rank anomalies (see Appendix Fig. B5 for all hotspots). Fig. 5b shows the correlation between the seasonal anomalies and decadal trends in the hotspots, only when the correlation is statistically significant and consistent in sign (see method section 2.2). From these matrices, it becomes evident that there are two regional clusters, where both the seasonal and decadal behavior is aligned. We define the two regions as follows; The US-Atlantic region (green box in Fig. 5b,c) consists of the eastern US Low, the Greenland High, and the North Atlantic Low; And the Eurasia region (yellow box in Fig. 5b,c) consists of the west Russian High, the Caspian Low, and the west China High.

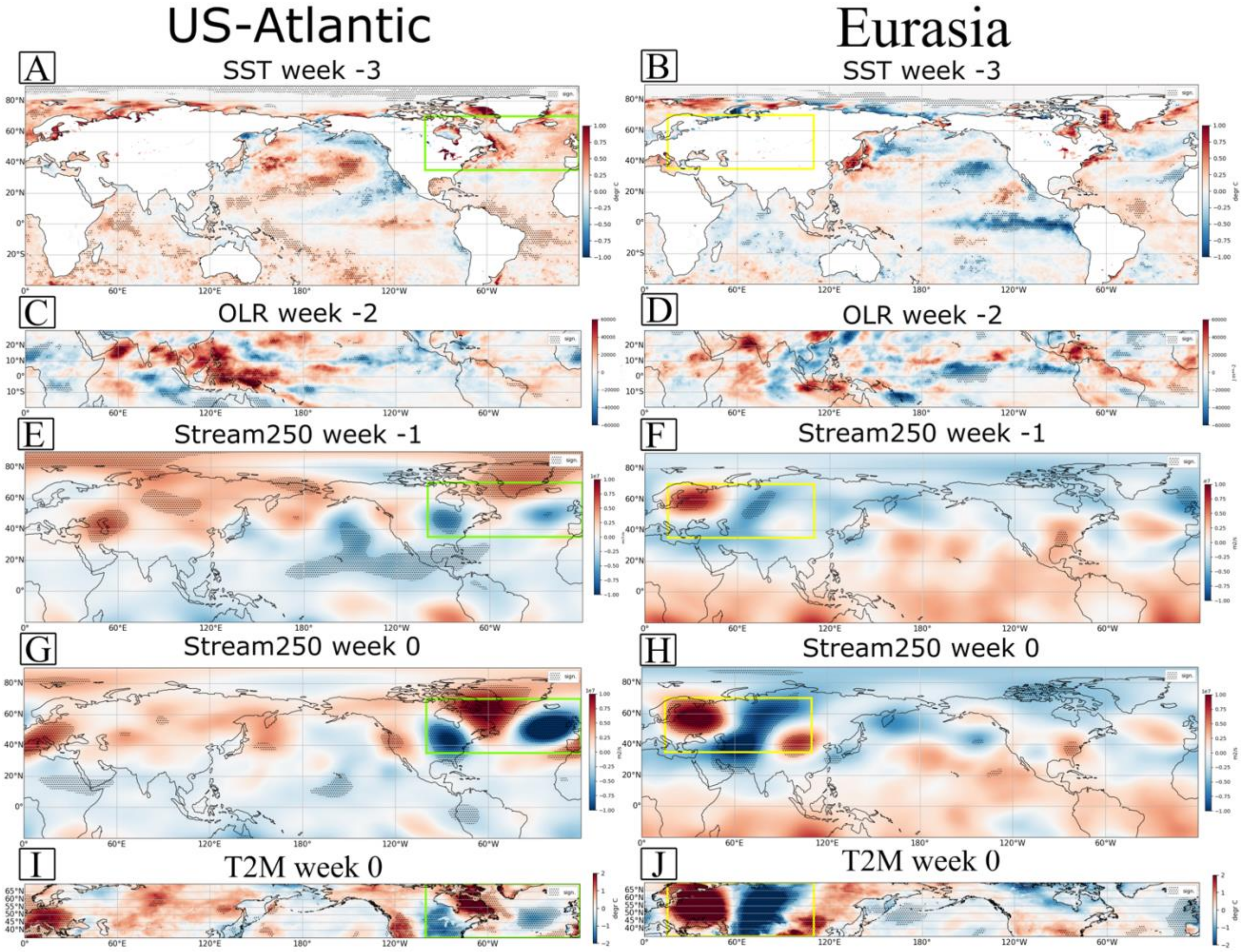

**Figure 6. Composites of positive matches on trend pattern for the two regions US-Atlantic (panels A,C,E,G,I) and Eurasia (panels B,D,F,H,J). A), B)** SST anomalies 3 weeks before in degrees C, **C), D)** OLR anomalies 2 weeks before in $Jm^{-2}$, **E), F),** streamfunction at 250hPa 1 week before in $m^2/s$, **G), H)** streamfunction at 250hPa during the week in $m^2/s$ and **I), J)** the T2M surface imprint during the week in degrees C. Stippling indicates statistical significance with a double sided t-test (Welch's) using p=0.05. Here we show the composites of the positive matches between the period of 1950-2014, to be able to compare to the CMIP6 model simulations.

### 3.4 Processes leading up to the patterns

Based on the above analyses (section 3.3), we can conclude that the US-Atlantic and Eurasia segments of the trend pattern are statistically behaving differently. To assess possible physical mechanisms that lead to the observed atmospheric anomalies, we use pattern matching to find weeks where the streamfunction anomalies look like the trend pattern in each region (see method section 2.3). Doing so, we identify the processes that are increasing in frequency over time. Here we analyse the anomalies in different variables leading up to the atmospheric configuration of interest. Figure 6 shows the composites of weekly mean anomalies of SSTs (- 3 weeks), OLR (-2 weeks), streamfunction (-1 and 0 weeks), and TAS (0 weeks), for the two different regions. The two regions show clear distinct patterns in all variables, suggesting that indeed different processes lead to these two different atmospheric configurations. The US-Atlantic pattern is preceded by statistically significant positive SST anomalies in the North Pacific with negative SST anomalies arcing around, resembling a negative PDO phase, and positive SST anomalies in the tropical pacific, resembling a positive ENSO phase (Fig. 6a). It furthermore shows statistically significant positive OLR anomalies, indicating reduced convective activity, which is bordered by significant negative OLR anomalies, resulting in a pronounced tripole-structure in the tropical west Pacific (Fig. 6c). The streamfunction anomalies (Fig. 6e,g) show wave-like structures up-stream the US-Atlantic region, which are associated with significant temperature anomalies at the surface (Fig. 6i), also beyond the defined regions. The Eurasian pattern shows contrasting SST anomalies to the US-Atlantic, with statistically significant negative SST anomalies in the North Pacific, indicating positive PDO phase, and strong negative SST anomalies in the equatorial Pacific, resembling a negative ENSO phase (Fig 6b). Additionally, the North Atlantic Ocean shows tripole structured anomalies as well, although not all of these are statistically significant. The OLR has less clear patterns, but indicates enhanced convection in the west Pacific (Fig. 6d), with especially strong convection in South East Asia. The streamfunction anomalies show a wave-like structure strengthening over the North Atlantic (Fig. 6f,h). Extremely high temperature anomalies are seen in Russia, with clear negative temperature anomalies east of that (Fig. 6j). Central China and the eastern US also show statistically significantly higher temperatures.

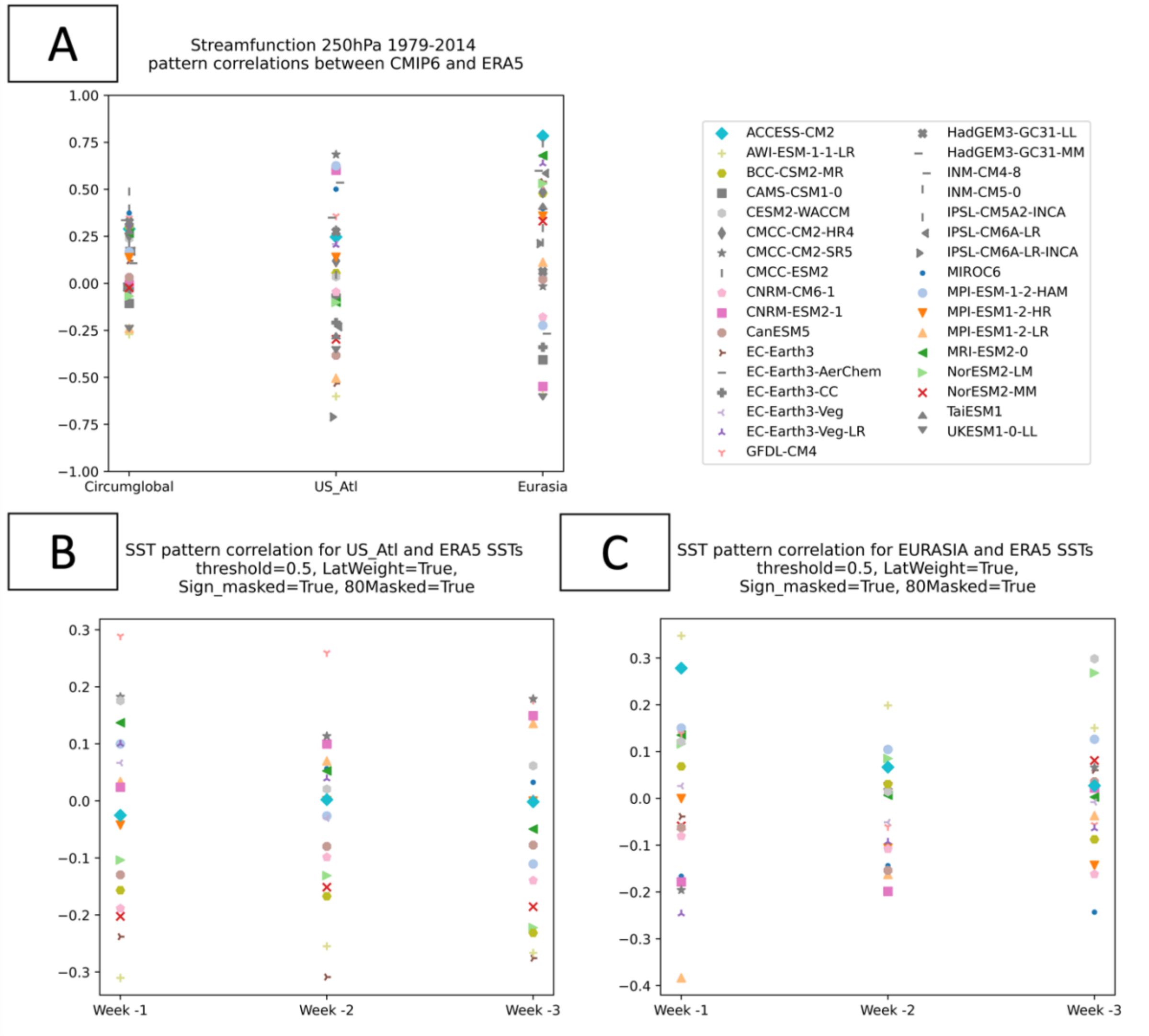

**Figure 7. Pattern correlations between CMIP6 and ERA5 a)** for Streamfunction 250 hPa trend; **b)** for US-Atlantic SST composites for different timelags; **c)** for Eurasia SST composites for different timelags.

### 3.5 Observations versus CMIP6

To assess whether CMIP6 model simulations accurately represent the same trend patterns, we first investigate the pattern correlation between the model's streamfunction trends and the observed ones, for different regions. Fig 7a shows that few models have similar trend patterns as ERA5 for the entire NH midlatitude region, with the maximum value of 0.47 for INM-CM5-0. For the individual trend pattern regions, we see the pattern correlation diverging, with some models showing an increase while others a decrease. For example, ACCESS-CM2 has a high pattern correlation (0.78) with the ERA5 trend pattern when considering only Eurasia, and CMCC-CM2-SR5 for US-Atlantic (0.69). This indicates that, for the individual regions, some climate models do capture the correct trend pattern. Appendix Figure B2 shows the trend patterns for those models with the highest correlations in the individual regions, and those with the highest and lowest correlation for the circumglobal region. Here, we see that even when the trend in a climate model simulation has a high correlation with ERA5, the pattern is still less pronounced than those observed in ERA5 (Fig. B2, Fig. B3). Which is in line with our previous findings, where we show that circulation trends are generally weaker in CMIP6 than in ERA5 (Fig. B4).

Secondly, we are interested to learn whether the CMIP6 models accurately represent the SST-to-wave pathways as found in ERA5. To investigate the pathways preceding the atmospheric configurations in the two regions, we repeat the composite analysis for the CMIP6 models, matching on the ERA5 regional trend patterns for the period 1979-2014 (see appendix Fig. B6 for the combined composite of all CMIP6 models). The super-composite of CMIP6 also shows a wave-like atmospheric circulation, with minor differences from ERA5. We then calculate the pattern correlation between the SST anomalies of each individual CMIP6 model and the SST anomalies as found in ERA5 (for the period 1979-2014), masking out the non-significant grid points and grid points above 80 degrees North, due to spurious significance values. We find that, overall, the pattern correlations are weak, with only very few reaching a correlation of ~0.3 (Fig. 7b-c). This indicates that for the same atmospheric wave patterns, the CMIP6 models do not have the same preceding SST anomalies as observed in ERA5.

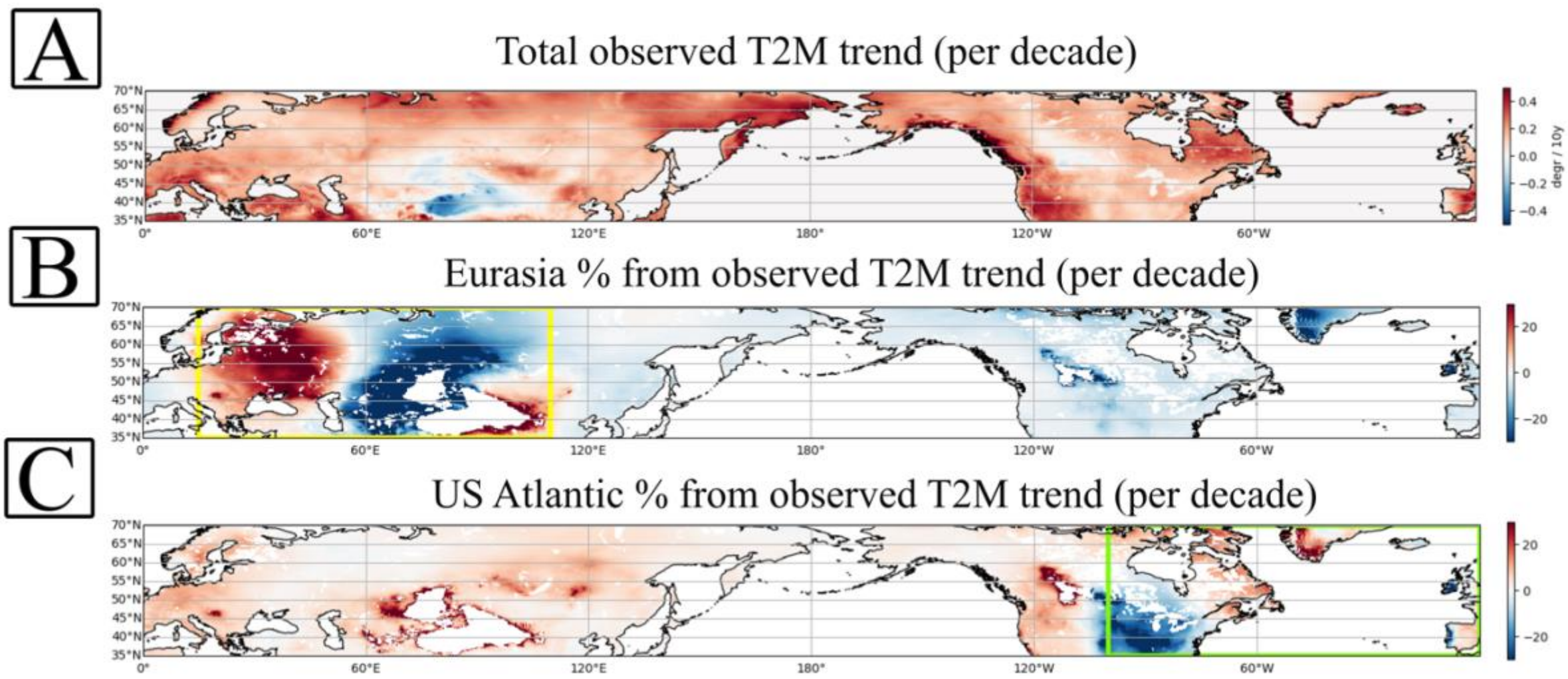

**Fig 8. A) The observed T2M trend and B,C) the contributions of the two different trend patterns.** The regions with very small trends (from panel A) are masked out to avoid spurious percentages in panels B and C.

### 3.6 Attributing T2M trends to circulation

Lastly, to assess to what extent the circulation trends in the two different regions contribute to the observed T2M trends, we train a simple regression model on the Global Mean Surface Temperature (GMST) and the pattern correlation with the circulation trend (see method section 2.4 for more details). We find that the changes in the US Atlantic trend pattern can explain on average 15% of the observed temperature trends in ERA5 in the US Atlantic region (Fig. 8C). Locally, even higher fractions of the temperature trend can be explained. The trend pattern is associated with higher temperatures in western US, Greenland, and Europe, which is in line with the results from the composite analysis (Fig. 6I). It furthermore offsets part of the positive temperature trend from the increase in GMST in the eastern US. Similarly, the Eurasia trend pattern explains on average 26% of the observed T2M trends in the Eurasian region (Fig 8B), with again much higher explained fractions locally. Similar to the US Atlantic model, the GMST trend and coefficients show a uniform temperature increase (Fig. B7), whereas the circulation trends show specific contributions to regional patterns.

## 4. Discussion

In this study, we find that the observed circulation trends in the upper-atmosphere consist of at least two separate regional patterns with a low-high-low over the US-Atlantic and a high-low-high over Eurasia. We show that these two atmospheric configurations are preceded by different SST and OLR anomalies and have significant temperature imprints at the surface too. The trend in these atmospheric configurations can explain on average 15% (US-Atlantic) and 26% (Eurasia) of regional temperature trends, as observed in ERA5. Lastly, CMIP6 simulations fail to accurately capture the atmospheric circulation trends and we show that this mismatch may be due to missing teleconnections from SSTs to the atmosphere.

### 4.1 Circulation and temperature trends

ERA5 observations show a marked set of changes in both surface and upper atmospheric circulation. We find that the upper-atmospheric circulation trends are at least partly responsible for the surface temperature trends in the two identified regions, US-Atlantic (15%) and Eurasia (26%). This is in line with previous research which shows that part of the temperature trend in western Europe and eastern US can be attributed to changing dynamics (Vautard et al., 2023, Singh et al., 2023). Since we show that the CMIP6 model simulations often fail to capture the dynamical trends, this could have important implications for future projections of surface temperature with global warming. If these projections are underestimated, temperatures could rise even faster than we expect. We know from previous research that at least for some regions the temperature trends are not captured well by climate models (e.g. Patterson 2023). One of those regions is western Europe (e.g. Vautard et al., 2023). As opposed to other regions, the positive temperature trend in western Europe is not reflected by a positive circulation trend (Fig. 2). In our results, we show that the temperature trends in western Europe are at least partly due to the increase in the US-Atlantic wave pattern which favors deep lows over the eastern Atlantic (D'Andrea et al., 2024). In the right position, these lows can cause southerly flows over western Europe bringing warm air from Spain, or even further south, towards north-western Europe causing hot extremes (D'Andrea et al., 2024). Furthermore, we show that such atmospheric configurations can lead to temperature anomalies in multiple locations at the same time, which sometimes last 2 weeks. This can have high impacts on e.g. the agricultural sector, as the wheat production breadbaskets are mostly located in the northern hemisphere midlatitudes (Kornhuber et al. 2020). Considering the steep increase in temperatures and the fact that for some of these regions they are outside the CMIP6 model range (e.g. Vautard et al., 2023), we argue that future research should focus on understanding this mismatch and the implications it has on long term temperature projections. The vital question remains whether this is internal variability or a response to climate change.

### 4.2 Physical mechanisms

Using correlation analysis, we show that the observed northern hemispheric wave-like trend pattern in boreal summer circulation consists of two distinct regions: A US-Atlantic pattern and a Eurasian pattern. These two regions are statistically not linked to each other and also have different preceding SST and OLR anomalies. This suggests that the observed circumglobal wave-like trend pattern in circulation could be a superposition of two continental scale wave-patterns, which are increasing in frequency. We find that the two wave-patterns are consistent in magnitude over time (Fig. B9), confirming that the trend pattern is indeed due to an increase in frequency. The decadal correlation matrices do show that some regions further up-/downstream of the two regional-patterns are also significantly correlated. This is in line with our composite analysis, where there are some significant circulation anomalies up/downstream of the pattern as well.

Previous research has explained the observed atmospheric circulation trends as a Rossby wave with wavenumber 5 (Sun et al. 2022; Teng et al. 2022). However, our results suggest that the wave-like trend pattern consists of two regional atmospheric waves. This is in line with some previous results from both Kim & Lee (2022) and Teng et al. (2022) where only parts of the circumglobal pattern can be explained with forcing experiments. First of all, Kim

& Lee (2022) show that they can excite the Eurasian circulation anomalies from latent-heating forcing in the North Sea and Caspian Sea. Secondly, Teng et al. (2022) show that the changes in eddy vorticity forcing in different regions (globally, North Pacific, and North Atlantic) can only explain parts of the observed trend pattern. We furthermore find that the Eurasia part of the wave-like trend pattern only emerges after 1979, while the parts of the US-Atlantic pattern is already evident before 1979. This time-dependence is further shown in our correlation-analysis, which indicates that although the trend pattern visually appears as a circumglobal wave, it is actually the result from the superposition of several regional processes. We therefore argue that future research and experiments should focus on understanding these circulation changes individually.

*4.2.1 Eurasia*

The Eurasia pattern is associated with SST anomalies resembling a combination of La Nina and positive PDO conditions and a tripole structure in the North Atlantic Ocean (NA). Dipole and tripole like anomalies are known to increase the baroclinicity and therefore can act as a Rossby wave source. The West North Pacific SSTs are closely related to ENSO, which have been linked to the Eurasia circulation pattern previously (Ding et al., 2011, O'Reilly et al., 2019). Similarly, Teng et al. (2022) find that the circumglobal trend pattern is linked to multi-decadal variability in SST patterns, notably the PDO and AMV. Indeed, research has shown that a wave over Eurasia can be due to SST and upper-level-convergence trends in the NWP (Sun et al., 2022) as well as increased synoptic eddy momentum transport (Teng et al. 2022). Next to the Pacific Ocean, research has also indicated that NA SST anomalies are important, as the NA can act as a bridge by strengthening waves originating in the WNP (Chen and Zhou 2018; Sun et al. 2022; van Straaten et al. 2023). For example, Sun et al. (2022) argue that Rossby waves coming from the WNP can lead to a cyclonic structure over the NA, after which wave energy propagates in a southward arch to central Asia. This is in line with O'Reilly et al. (2019), who link both cyclonic NA and Eurasia circulation anomalies to SSTs in the pacific. Conversely, Nie et al. (2024) find that the Eurasia circulation pattern is preceded by anticyclonic conditions over the NA region, which is in line with previous research (Sun et al. 2008; Chen and Zhou 2018). Lastly, Kim and Lee (2022) find that the Eurasia circulation pattern, and associated temperature, is associated with increased latent heating in the NA in combination with suppressed latent heating in the Caspian Sea. Apart from SST, soil moisture depletion has also shown to influence Rossby waves (Teng et al., 2022). From our result, we hypothesise that the wave activity is likely a combination from enhanced activity in the Pacific and downstream amplification by NA SST anomalies, forming the final Rossby wave that leads to the Eurasia circulation pattern.

### *4.2.2 US-Atlantic*

The US-Atlantic pattern is associated with SST anomalies representing a combination of El Nino and negative PDO conditions. In contrast to the Eurasia composites, there does not seem to be notable SST anomaly patterns in the NA region. However, both the tropical and northern Pacific have previously been linked to atmospheric Rossby waves (Vijverberg and Coumou 2022). The SST anomalies in our findings resemble positive ENSO and negative PDO conditions, which is a relatively rare combination (Feng and Tung, 2020). Experiments changing the eddy momentum fluxes in the North Pacific did not reveal the trend pattern as observed (Teng et al. 2022), suggesting this region alone cannot explain the circulation trends in the US-Atlantic region. However, experiments changing the diabatic heating in the tropical Pacific does result in the trend pattern over the US-Atlantic (Sun et al. 2022). Furthermore, CMIP6 models fail to capture the increase in the Greenland blocking (Davidini & D'Andrea, 2020; Maddison et al., 2024), which is partly driven by SST, Sea Ice and/or anthropogenic aerosols (Maddison et al., 2024). Aerosols are known to be able to influence atmospheric circulation patterns by modifying the distribution of solar radiation (Akinyoola et al., 2024). Furthermore, climate models often have biases in aerosol representations, which can lead to changes in circulation responses (Brown et al., 2021, Chemke & Coumou, 2024). Future research is needed to indicate what is causing the increase in the US-Atlantic wave pattern, and whether this is due to anthropogenic activities.

## 4.3 Observations and CMIP6

We find that CMIP6 models do not exhibit the correct trend pattern nor the right magnitude of atmospheric circulation trends, compared to ERA5 (Fig. 7, Fig. B2, Fig. B3, and Fig. B4). Furthermore, they do not exhibit the same precursor-anomalies that lead to the respective atmospheric configurations as seen in observations. Even though the pattern correlations between the observed trend pattern and the CMIP6 models show higher correlations for the two individual regions (US-Atlantic and Eurasia), as compared to the entire circumglobal midlatitudes, most models do not show the right pattern. We have also shown that even when certain models capture parts of the trend pattern, the magnitude of the trends in the observations are much higher (Fig. 3e, Fig. B3, B4).

The underestimation of the magnitude of circulation trends, as compared to observations, is in line with previous research (Singh et al. 2023; Vautard et al. 2023; D'Andrea et al. 2024). This mismatch is potentially linked to the signal-to-noise paradox of climate models, which argues that the climate models' response to external forcing and boundary conditions is too weak (Scaife & Smith, 2018). A potential reason for the weak response of climate models lies in a lack of extratropical ocean–atmosphere coupling (Scaife & Smith, 2018). Our results support this hypothesis, as we find the SST anomalies prior to the atmospheric pattern to be much weaker in CMIP6 than ERA5. This suggests that CMIP6 models struggle with reproducing boreal summer teleconnections, as shown in previous research too (Di Capua et al., 2023). Specifically, climate models may lack the correct teleconnection from the tropical Pacific to Rossby wave activity in the atmosphere (Jong et al. 2021; Sun et al. 2022), and thereby also failing to accurately predict surface imprints (van Straaten et al. 2023). A possible explanation for the weak teleconnection in climate models is biases in the extratropical jet (O'Reilley et al., 2019). A potential missing teleconnection or weaker response is worrying, as it could have large implications for regional climate projections if the response is weaker than in the real world.

It is vital to better understand the reasons for the mismatch in trends between models and observations, to be able to interpret the observed accelerated trends more accurately. Currently, it is still unclear to what degree the observed accelerated trends are due to natural variability or to what extent it is due to external forcing (Shaw et al., 2024, *preprint*). Liang et al. (2024), find that a substantial fraction of the underestimation of the global mean surface temperature is coming from internal variability in SSTs in the  Pacific. Thus, a systemic bias in Pacific SSTs in climate models can have large implications for climate projections (Liang et al., 2024). Our results provide further evidence that climate models are not accurately capturing teleconnection patterns from the Tropical Pacific Ocean to Northern Hemispheric midlatitude circulation in summer, and its associated surface imprints. However, future research should not be limited to just SSTs, as variables like soil-moisture (Teng et al., 2022; Kim and Lee, 2022), latent-heating (Kim and Lee, 2022), aerosols (Maddison et al., 2024), and snow cover (Preece et al., 2023; Maddison et al. 2024) have been linked to the circulation trends and mismatches in CMIP6 as well.

## 5. Conclusion

In this research we use various data-driven techniques to understand the distinct circulation trend pattern in the boreal summer in the northern hemispheric midlatitudes, a pattern which is not captured by the CMIP6 multi-model-mean and which lies outside of the CMIP6 model range for many of the individual hotspots. The observed trends in the northern hemispheric midlatitudes consist of a series of hotspot regions, with both positive and negative trends in circulation as well as connected surface trends in temperature. In our research, we show that the observed trend pattern consists of at least two separate regional patterns: the US-Atlantic and Eurasia regions. We show that these two regional circulation trend patterns can explain 15% (US-Atlantic) and 26% (Eurasia) of regional temperature trends in ERA5. Additionally, we find that the pattern correlations between the observed trends and the CMIP6 model simulations increase for the individual regions, compared to the circumglobal trend pattern, although they are still weaker than observed. The increased pattern correlation for the individual regions indicates that some climate models may accurately capture part of the observed trends, whether forced or due to internal variability.

We find that the local wave-like circulation patterns in both regions (US-Atlantic and Eurasia) are preceded by a distinct set of SST and OLR anomalies in the Tropical Pacific, North Pacific, and/or the North Atlantic Ocean. Furthermore, the wave-like circulation patterns are associated with strong T2M imprints at the surface on multiple locations. We show that for the same atmospheric circulation patterns, CMIP6 models do not have the same preceding SST patterns as found in ERA5. This suggests that the reason that the CMIP6 models do not capture the observed trend pattern is due to missing teleconnections originating from SST anomalies. To better understand the mismatch between climate models and observations, further research should focus on understanding the physical pathway for each of these regional waves, why the models fail to accurately model this physical pathway, and lastly how this influences short term predictions and long term climate projections.

*Acknowledgments.*

We thank the editor and reviewers for their time, questions, and remarks. This study was partly supported by the European Union’s Horizon 2020 research and innovation programme under Grant Agreement 101003469 (XAIDA project). We thank the maintainers and funders of the BAZIS cluster at VU Amsterdam for computational resources.

*Data Availability Statement.*

ERA5, CMIP6, and the NASA GMST data are publicly available. ERA5 data (Hersbach et al. 2020) were downloaded from the Copernicus Climate Change Service (C3S) Climate Data Store. NASA GMST data were downloaded from https://climate.nasa.gov/vital-signs/global-temperature/?intent=121 (accessed 29 oct 2024). The code will be made available on GitHub upon publishing [*link will be added*].

## APPENDIX

### Appendix A: pre-processing

Appendix A consists of additional information and explanation on the pre-processing steps of this research. Figure A1 shows the relative T2M trends for the midlatitudes, Figure A2 shows the northern hemispheric variability to ENSO, Figure A3 includes assessment of the rank anomaly method on a Gaussian toy model, and Table A1 includes the CMIP6 model names used in this research.

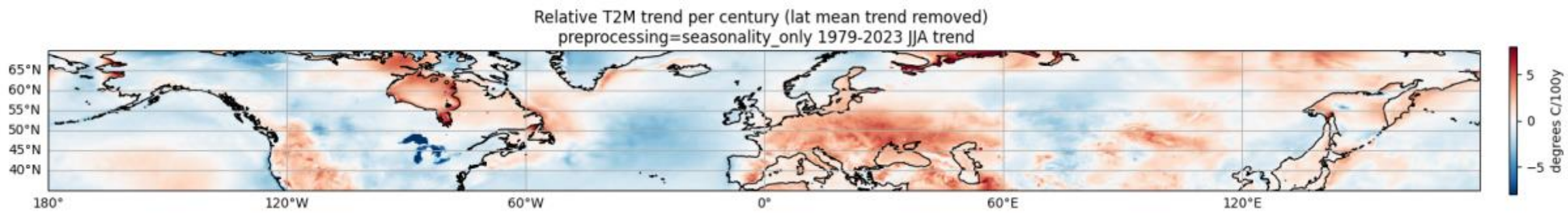

**Figure A1. Relative T2M trends in the midlatitude.** For each latitude, the mean latitudinal trend is removed to correct for the polar-amplification signal.

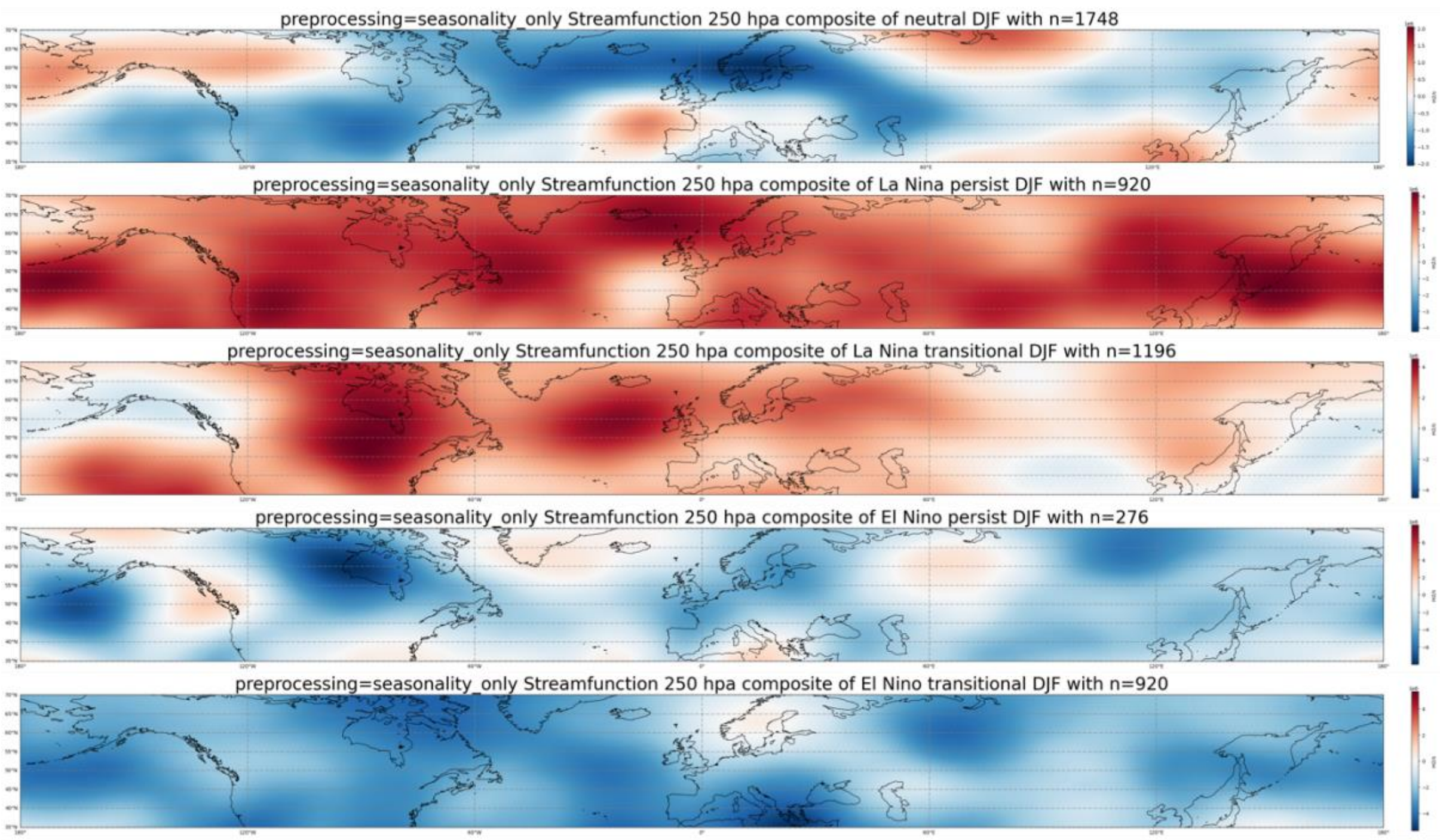

**Figure A2. Northern Hemispheric modes of variability connected to ENSO.** Separating the years based on their ENSO phase, we composite all those summers and find strong differences.

<table>
<tr><th>Settings</th><th>Visualisation without noise of mean and std over time</th><th>Random selected (fake!) years and the rank anomalies</th></tr>
<tr><td>Shift in Mean is False, shift in spread is True</td><td>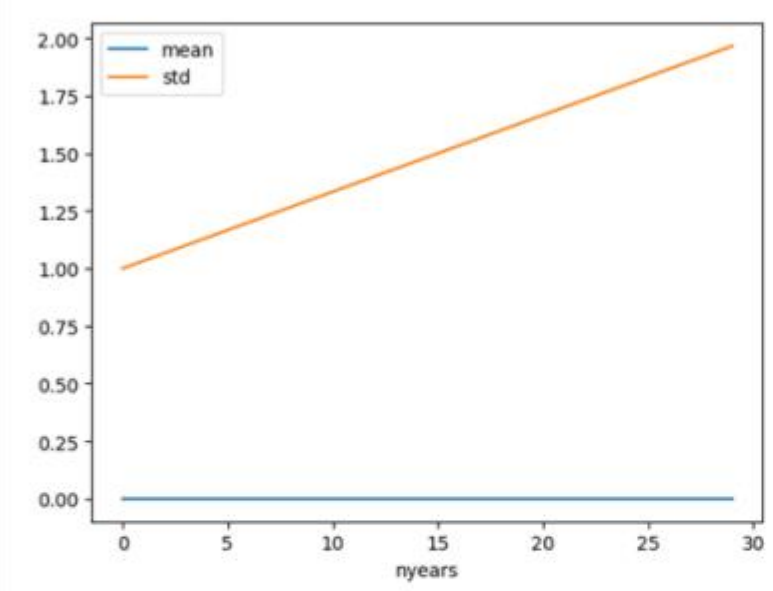
</td><td>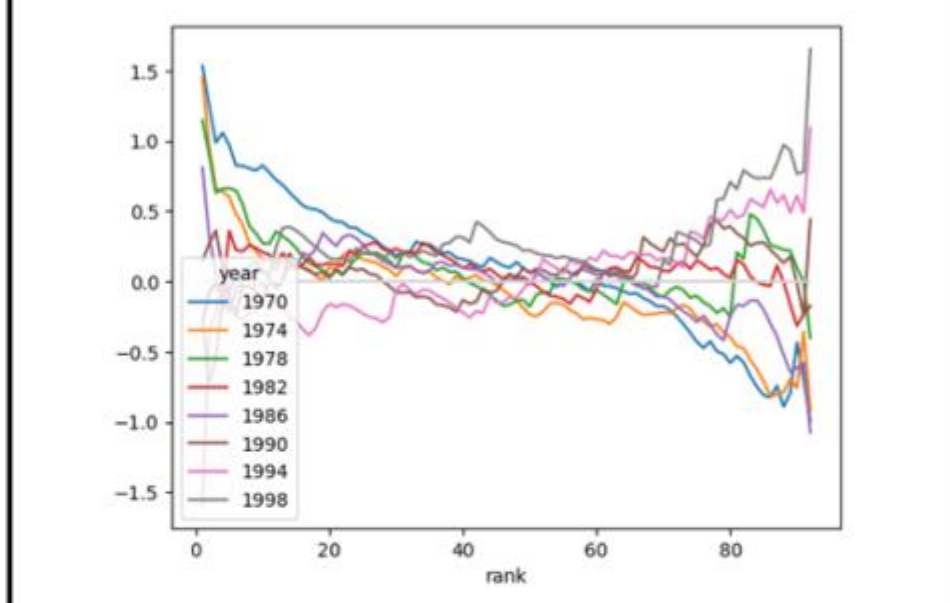
</td></tr>
<tr><td>Shift in Mean is True, shift in spread is False</td><td>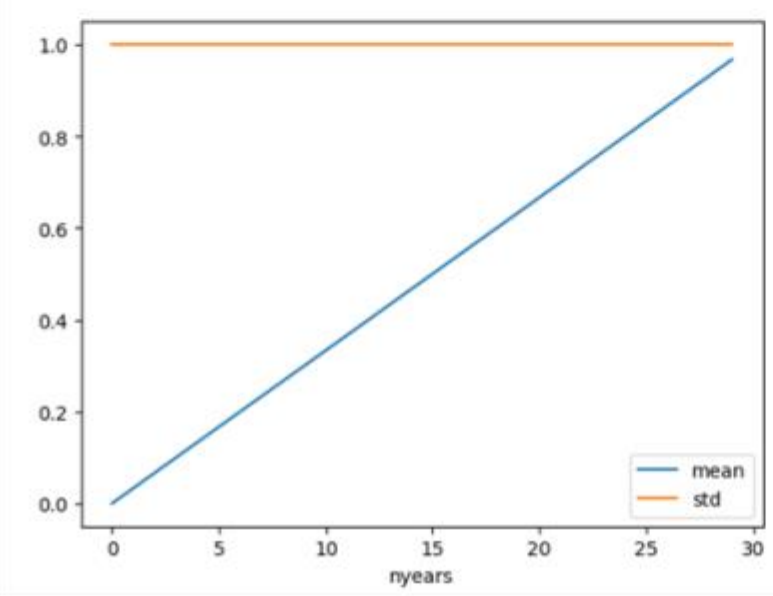
</td><td>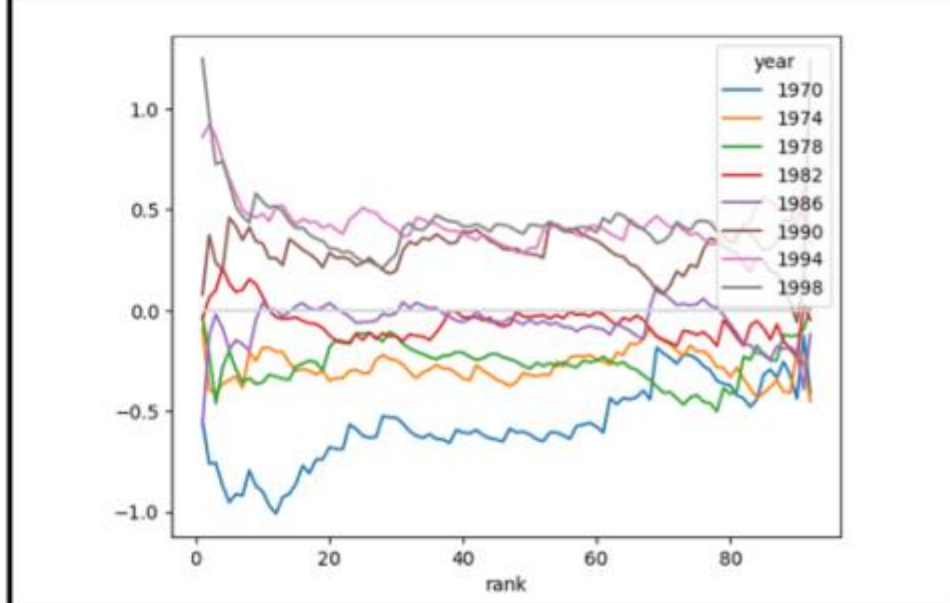
</td></tr>
<tr><td>Shift in both mean and spread</td><td>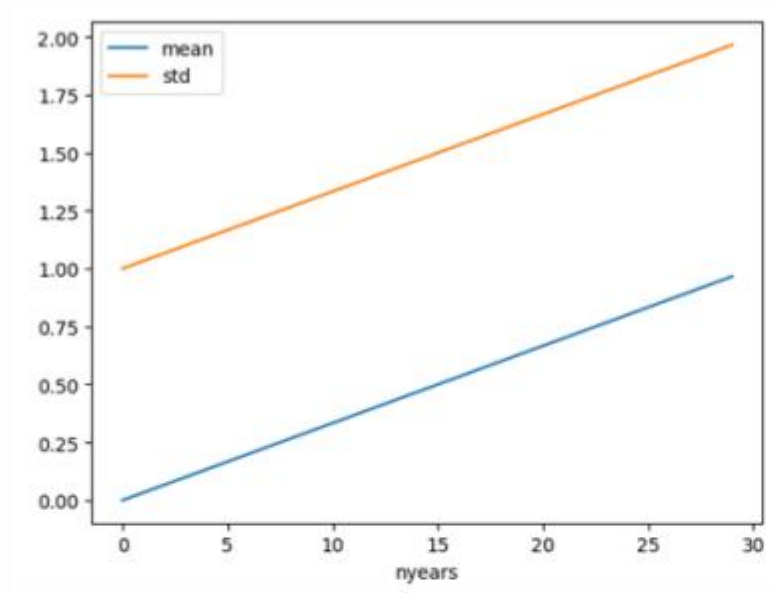
</td><td>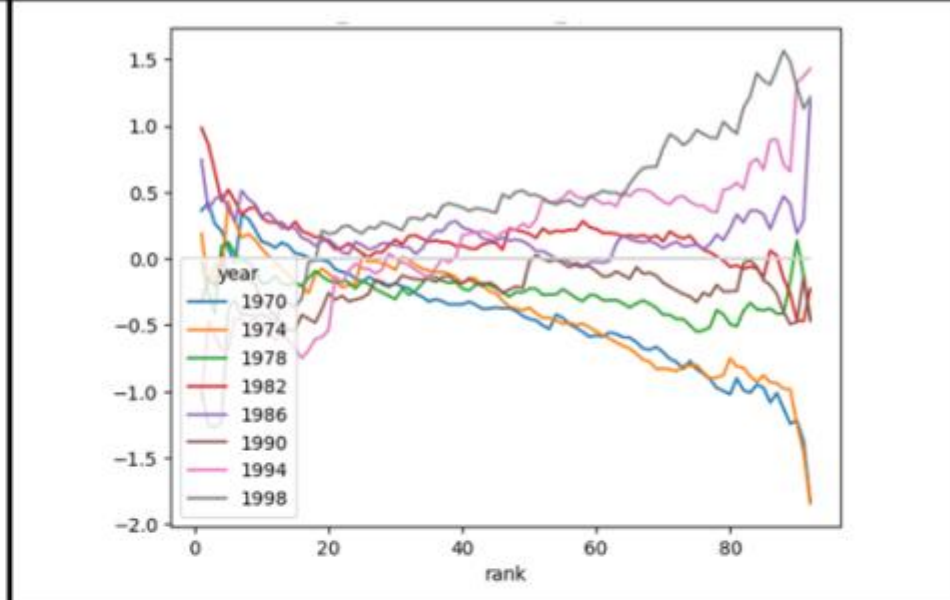
</td></tr>
<tr><td>No shifts</td><td>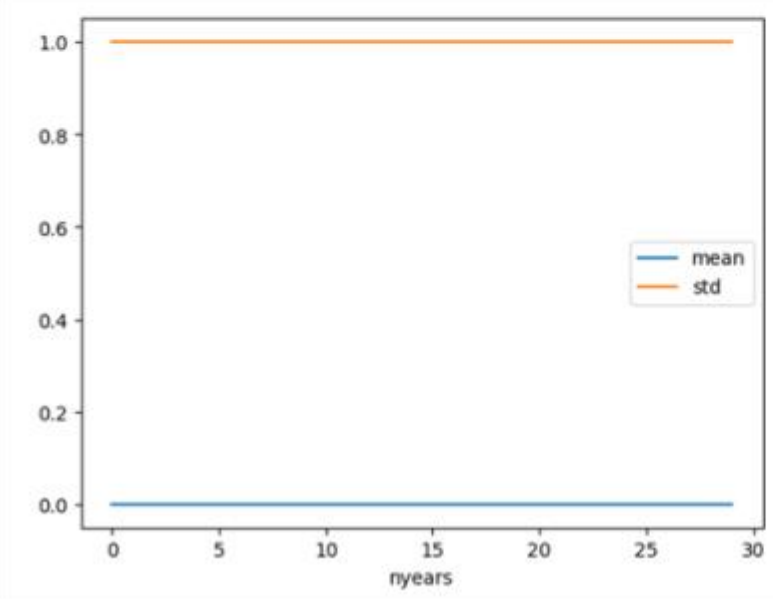
</td><td>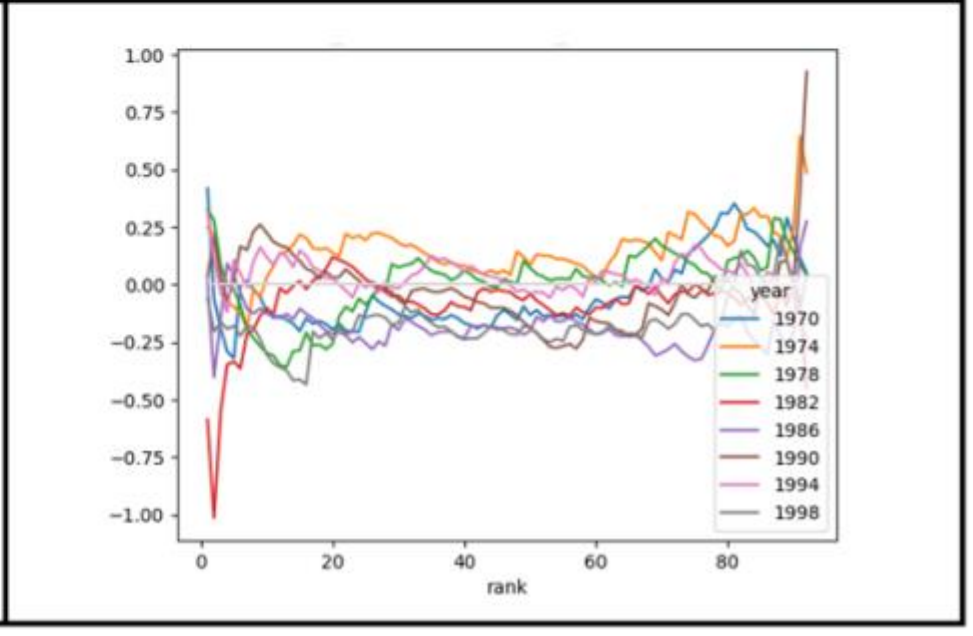
</td></tr>
</table>

**Figure A3 Gaussian Toy model and rank anomalies.** When assessing trends in the different parts of the rank anomalies (low, bulk, and high), we want to ensure we correctly interpret the changes. Therefore, we investigate what happens to the rank anomalies with different changes in the Gaussian distribution of our toy model: shift in mean = True/False, and shift in spread = True/False. The rank anomalies change as we would expect for all scenarios. When there's no shift, the years are randomly distributed around the mean anomaly line, with outliers everywhere. When there's a shift in only the mean, you see the later years shifting to higher anomalies (due to the positive shift). When there's a shift in spread, you see an increase in outliers at the tails of the rank anomalies.

**Table A1 cmip6 models used in this research.** CMIP6 Models used for streamfunction 250 hPa, in bold those we used for the SSTs comparison.

| | |
|---|---|
| **1.** | **ACCESS-CM2** |
| **2.** | **AWI-ESM-1-1-LR** |
| **3.** | **BCC-CSM2-MR** |
| 4. | CAMS-CSM1-0 |
| **5.** | **CESM2-WACCM** |
| 6. | CMCC-CM2-HR4 |
| **7.** | **CMCC-CM2-SR5** |
| 8. | CMCC-ESM2 |
| **9.** | **CNRM-CM6-1** |
| **10.** | **CNRM-ESM2-1** |
| **11.** | **CanESM5** |
| **12.** | **EC-Earth3** |
| 13. | EC-Earth3-AerChem |
| 14. | EC-Earth3-CC |
| **15.** | **EC-Earth3-Veg** |
| **16.** | **EC-Earth3-Veg-LR** |
| **17.** | **GFDL-CM4** |
| 18. | HadGEM3-GC31-LL |
| 19. | HadGEM3-GC31-MM |
| 20. | INM-CM4-8 |
| 21. | INM-CM5-0 |
| 22. | IPSL-CM5A2-INCA |
| 23. | IPSL-CM6A-LR |
| 24. | IPSL-CM6A-LR-INCA |
| **25.** | **MIROC6** |
| **26.** | **MPI-ESM-1-2-HAM** |
| **27.** | **MPI-ESM1-2-HR** |
| **28.** | **MPI-ESM1-2-LR** |
| **29.** | **MRI-ESM2-0** |
| **30.** | **NorESM2-LM** |
| **31.** | **NorESM2-MM** |
| 32. | TaiESM1 |
| 33. | UKESM1-0-LL |

## Appendix B: results

Appendix B consist of additional figures to compliment the result section of this research. Figure B1 shows the difference between the overall trend in circulation and the trend of the different parts of the distribution, Figure B2 shows some trend plots of CMIP6 models, Figure B3 shows Fig. 3E but with models labeled, Figure B4 shows the rank histogram for ERA5 versus CMIP6 trends, Figure B5 shows the decomposed timeseries for all hotspots, Figure B6 shows the super-composites for the CMIP6 models, Figure B7 shows the coefficients of the linear regression models used for the T2M attribution, Figure B8 offers a polar projection of the circulation trends, and Figure B9 shows the streamfunction 250hPa composites for two different time periods (as Fig. 6).

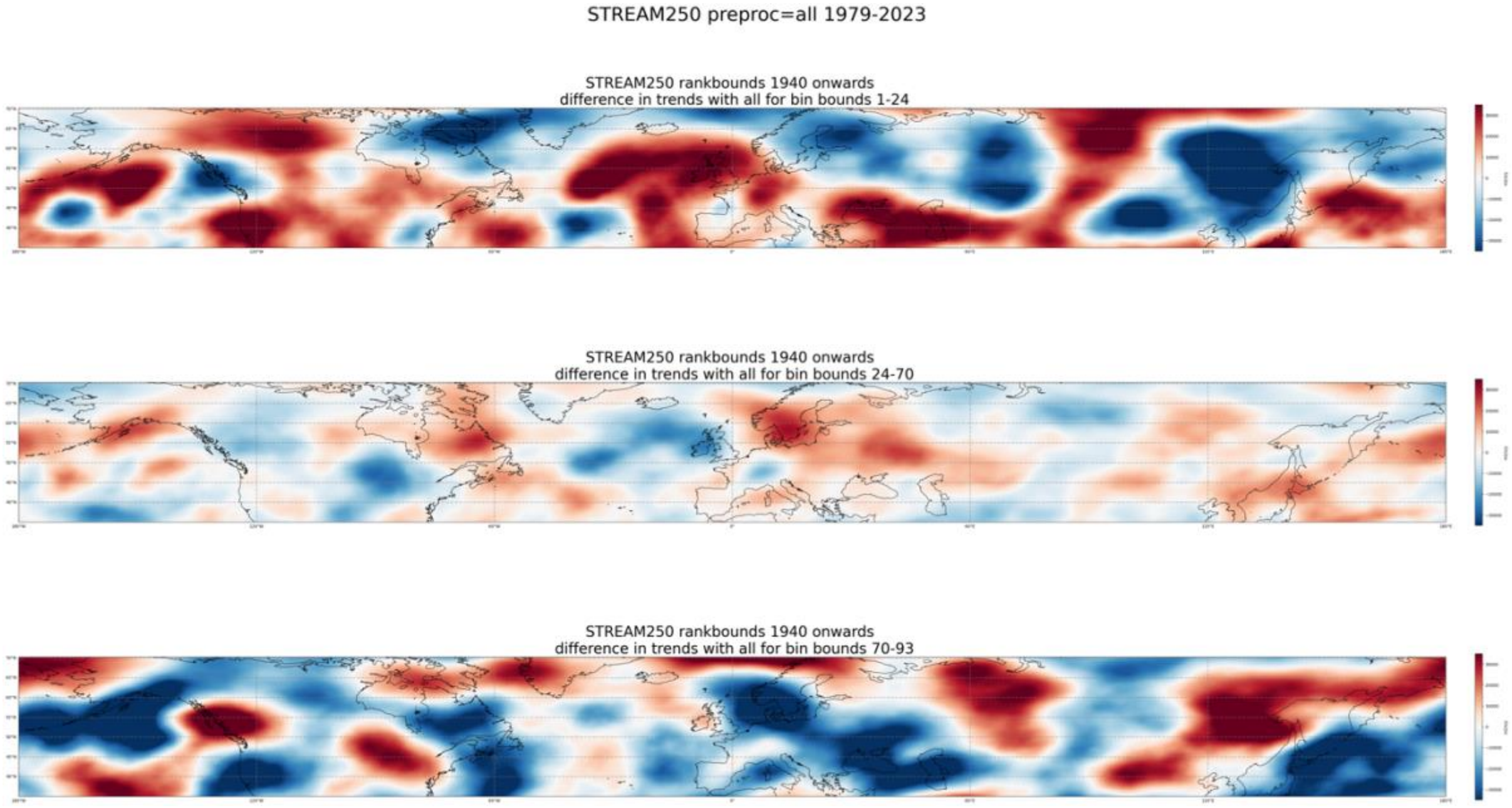

**Figure B1. Difference between trend pattern of the entire distribution and the trend of the low-, bulk-, and high-parts of the distribution.**

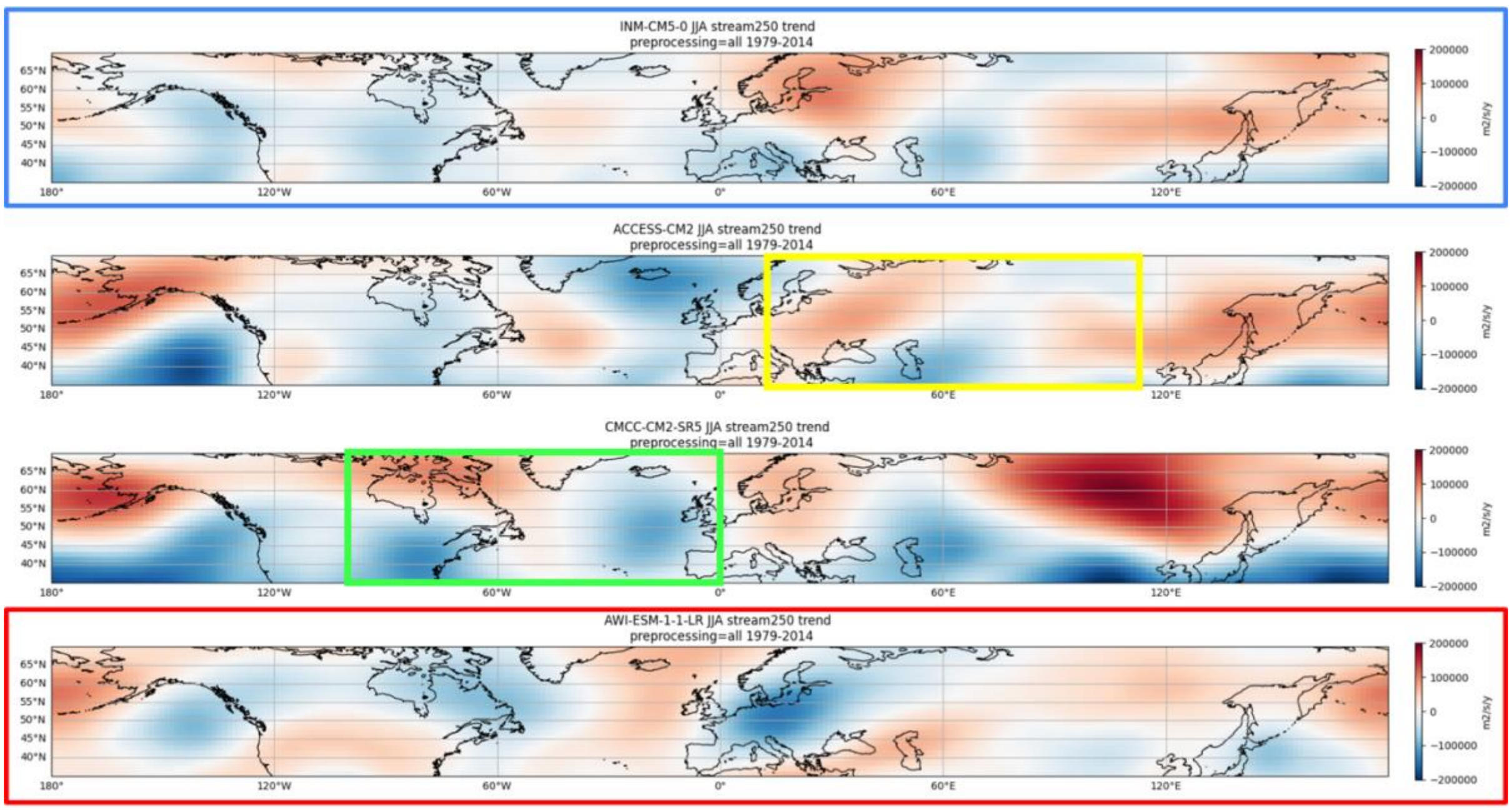

**Figure B2. Trendplots of streamfunction 250hPa,** for four climate models which show the highest pattern-matching with ERA5 for circumglobal (top), US-Atlantic and Eurasia (middle), and the lowest match with ERA5 circumglobal (bottom).

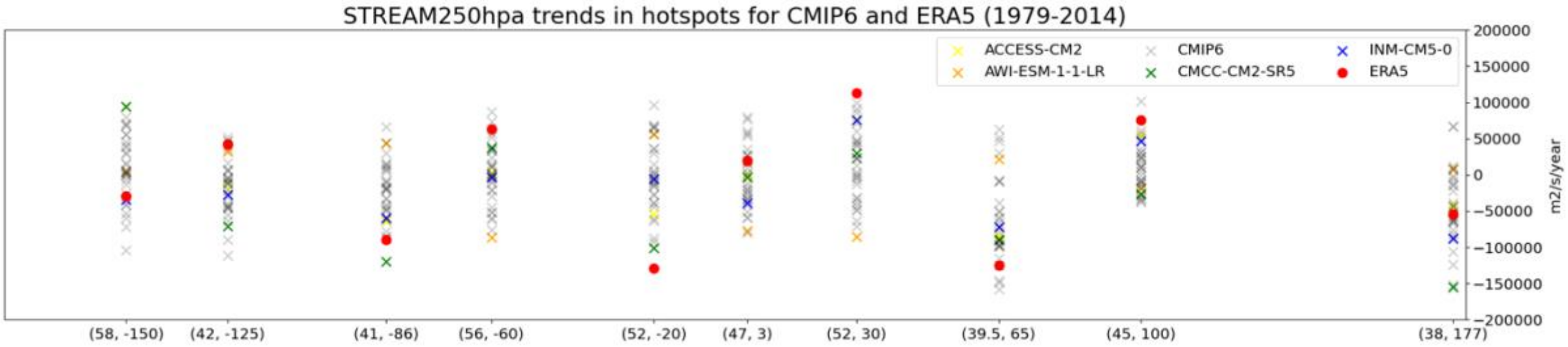

**Figure B3. Similar to Fig. 3E, but with models from Fig. B2 highlighted.**

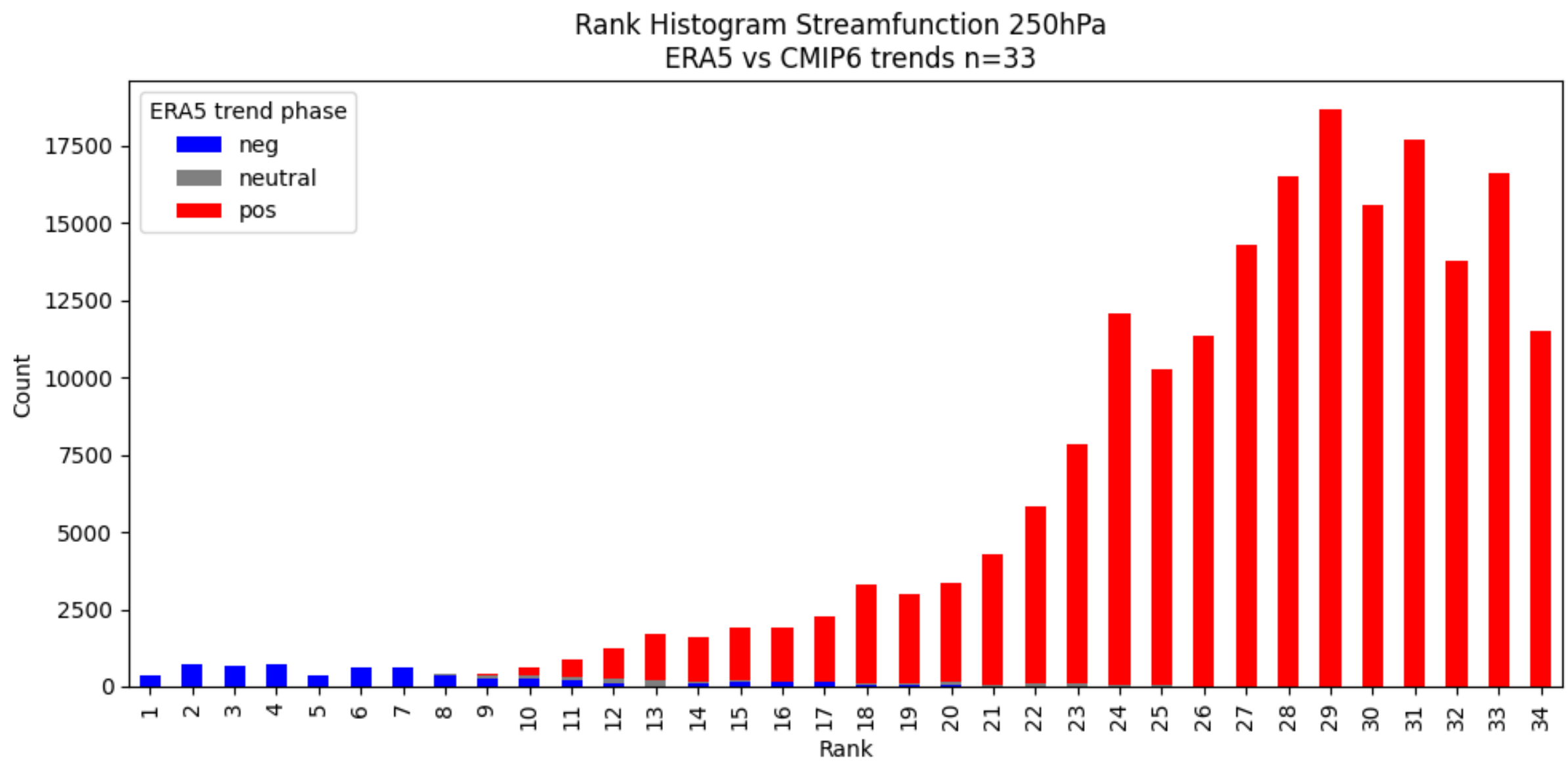

**Figure B4. Rank histogram of streamfunction 250 hPa for the midlatitudes.** For each gridpoint we calculate the rank that the observed ERA5 trend has in comparison to the 33 CMIP6 model simulations (e.g. rank 1 means all models have higher trends and rank 34 means all models have lower trends). We color each gridpoint depending on the sign of the ERA5 trend with three phases: the negative phase (<= -5000 m2/s), the neutral phase (in between), and the positive phase (>= 5000 m2/s).

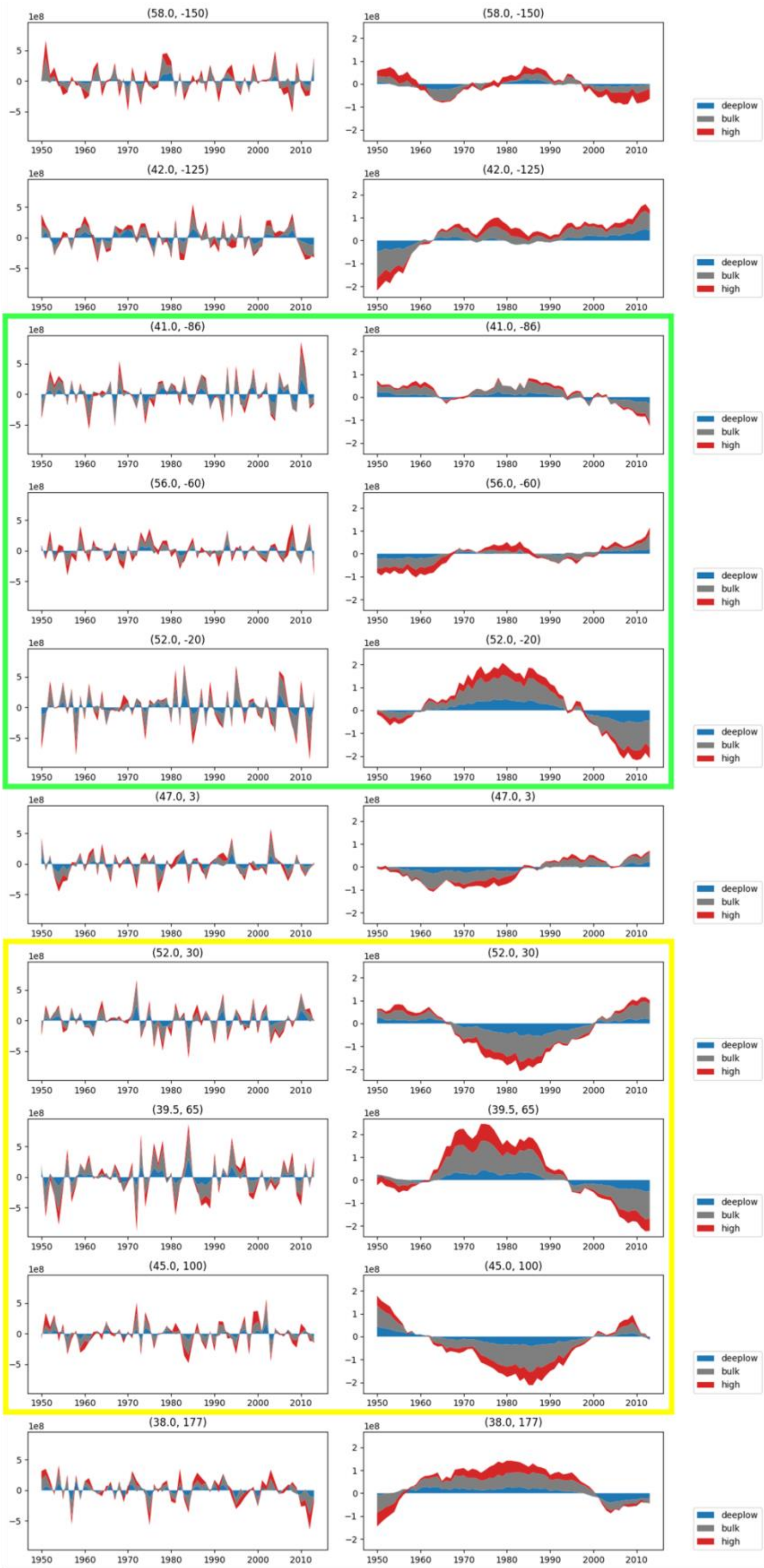

**Figure B5. Rank anomaly timeseries per hotspot**

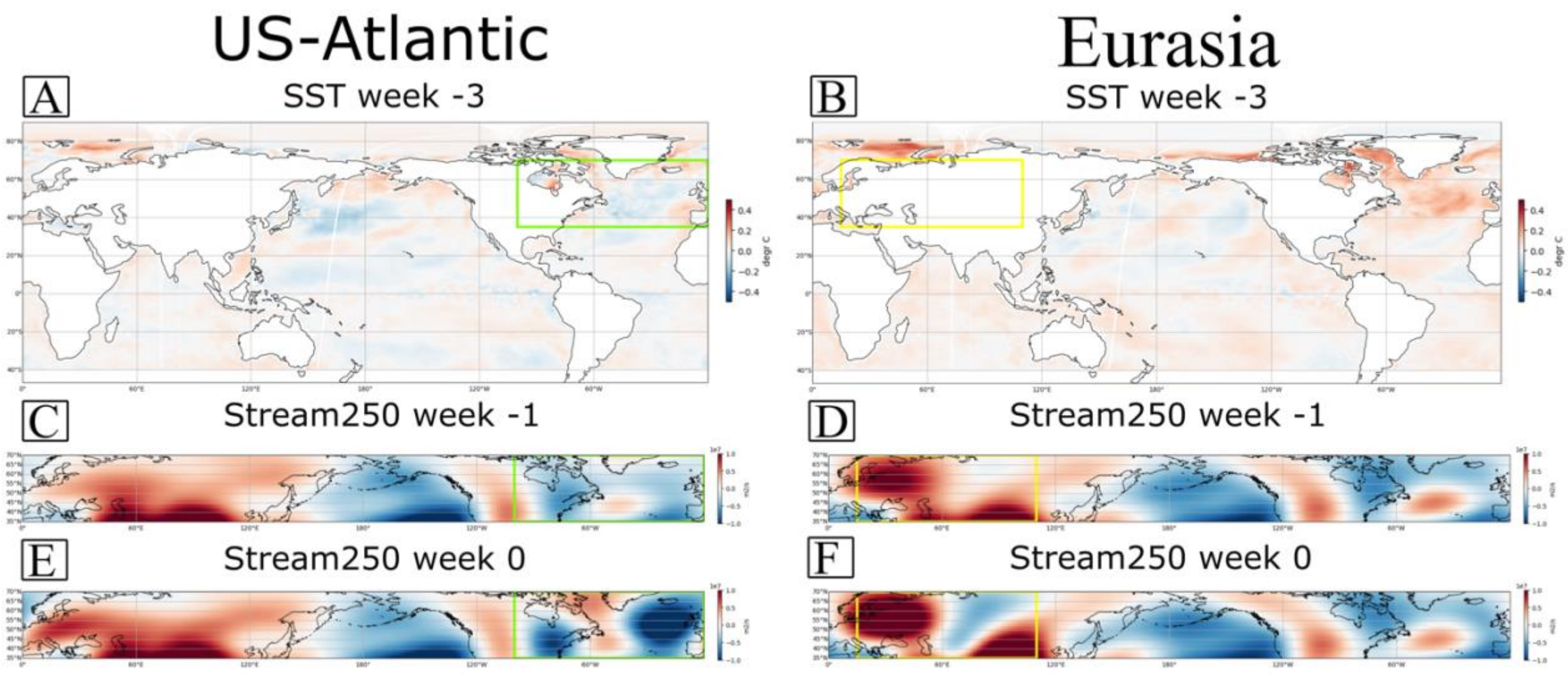

**Figure B6. Composites of (n=19) CMIP6 models.** Similar to figure 5, but then for all weeks from all CMIP6 models that have SST data, see table A1. Note that the scale bar for the SST plots in CMIP6 (-0.5, 0.5) is different from the scale bar of ERA5 (-1,1), due to the smaller values in the CMIP6 plots.

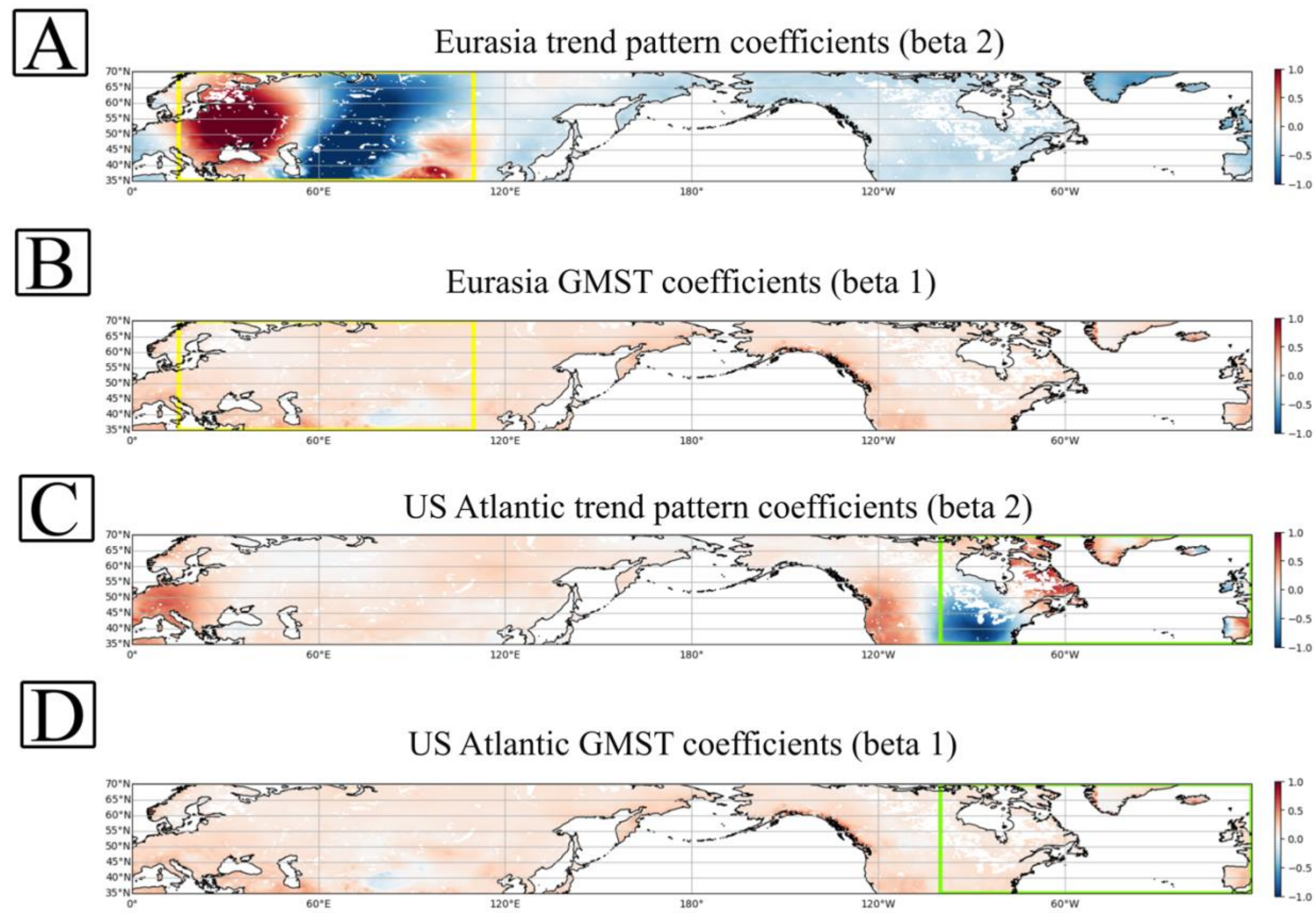

**Figure B7. The coefficients of the linear model predicting grid point T2M for Eurasia (A,B) and US Atlantic (C,D).** For details of the model see method section 2.4.

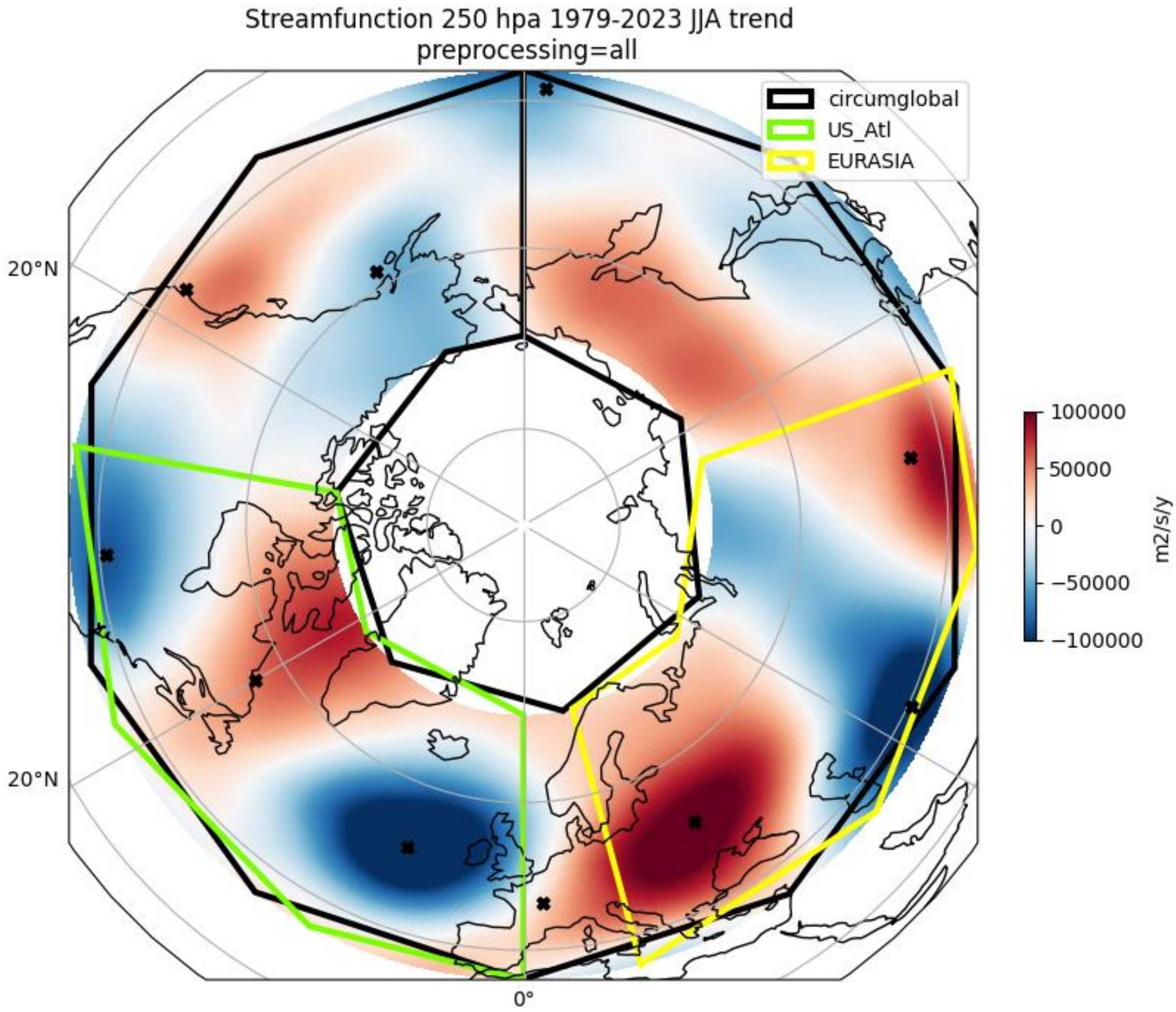

**Figure B8. Circumglobal trend pattern with polar projection.** Same as Fig 3C, but with different projection.

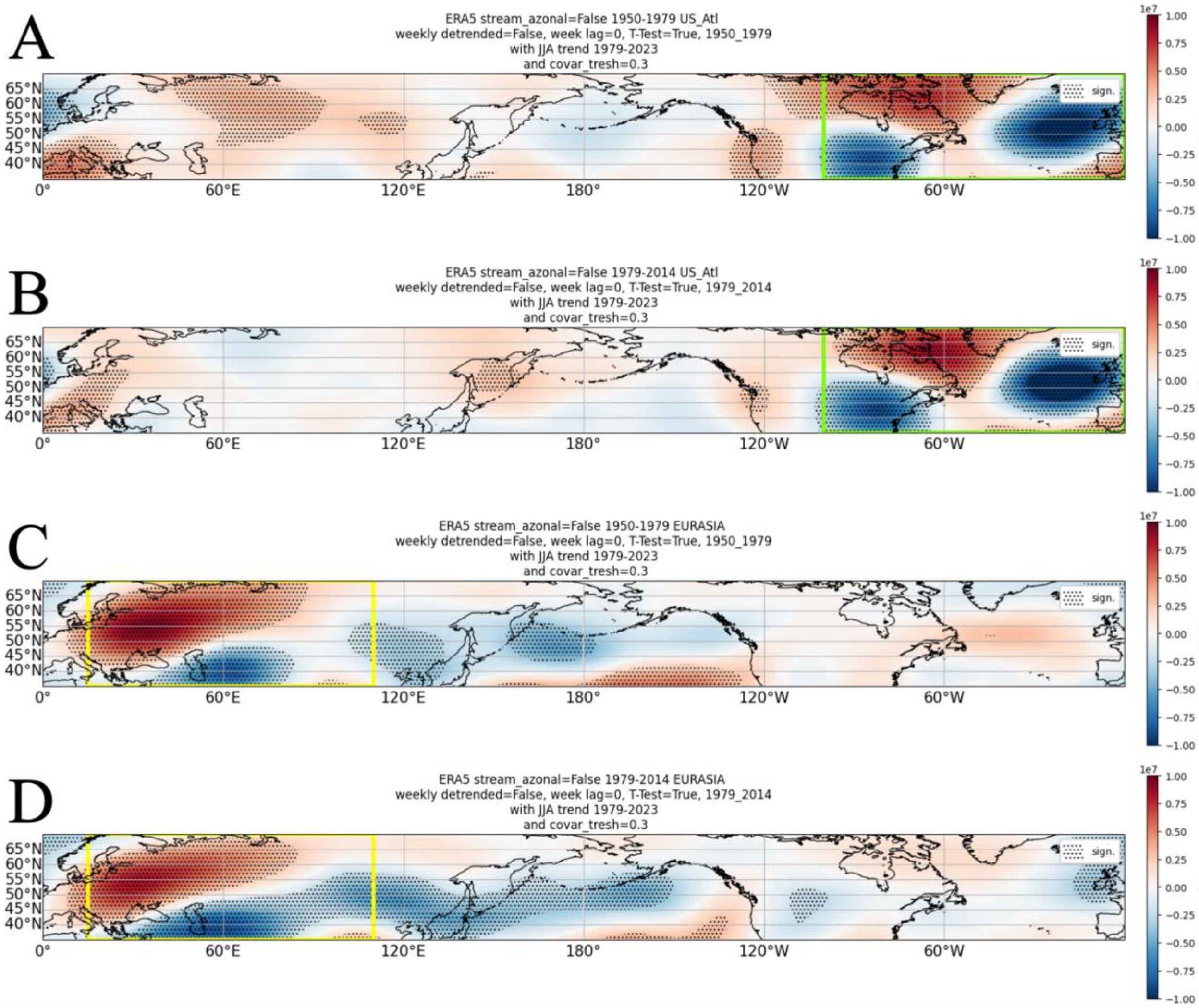

**Figure B9. Streamfunction 250 hPa composites for two different time periods,** as Figure 6 (with threshold of 0.3 to increase number of positive matches). A,B for the US-Atlantic, and C,D for Eurasia. A and C are from 1940-1979, and B and D for 1979-2014.